\documentclass[a4paper,fleqn]{article}

\usepackage{geometry}

\usepackage[british]{babel}
\usepackage{amsthm, amssymb, mathtools, amsmath}
\usepackage{color}
\usepackage{amsbsy}
\usepackage{enumerate}
\usepackage{mdwlist}
\usepackage{bigints}%\bigintssss, \bigintsss, \bigintss, \bigints, and \bigint

\usepackage{enumitem}

\usepackage[authoryear, round]{natbib}

\DeclareMathOperator{\N}{\mathbb{N}}
\DeclareMathOperator{\R}{\mathbb{R}}

\newcommand\restrict[1]{\raisebox{-.5ex}{$|$}_{#1}}

\newcommand{\B}[1]{\boldsymbol{#1}}

\newtheorem{remark}{Remark}
\newtheorem*{Observation}{Observation}

\usepackage{graphicx}
\usepackage{subfigure}
\usepackage{float}

\usepackage[normalem]{ulem} %\sout{text} strike out text

\usepackage[labelfont=bf,font=small,up]{caption}

\usepackage{lscape} %pages in landscape format \begin{landscape} ... \end{landscape}

\usepackage{url, verbatim}

\numberwithin{equation}{section}

\usepackage{etoolbox}
\makeatletter
\patchcmd{\maketitle}{\@fnsymbol}{\@alph}{}{}  % Footnote numbers from symbols to small letters
\makeatother

\title{Dangerous connections: on binding site models of infectious disease dynamics}
\date{\today}
\author{KaYin Leung\thanks{k.y.leung@uu.nl, Mathematical Institute, Utrecht University, The Netherlands}\ \ \thanks{Julius Center for Primary Care and Health Sciences, University Medical Center Utrecht, Utrecht, The Netherlands } \and Odo Diekmann\footnotemark[1]}

\begin{document}
\maketitle

\begin{abstract}
We formulate models for the spread of infection on networks that are amenable to analysis in the large population limit. We distinguish three different levels: (1) binding sites, (2) individuals, and (3) the population. In the tradition of Physiologically Structured Population Models, the formulation starts on the individual level. Influences from the `outside world' on an individual are captured by environmental variables. These environmental variables are population level quantities. A key characteristic of the network models is that individuals can be decomposed into a number of conditionally independent components: each individual has a fixed number of `binding sites' for partners. The Markov chain dynamics of binding sites are described by only a few equations. In particular, individual-level probabilities are obtained from binding-site-level probabilities by combinatorics while population-level quantities are obtained by averaging over individuals in the population. Thus we are able to characterize population-level epidemiological quantities, such as $R_0$, $r$, the final size, and the endemic equilibrium, in terms of the corresponding variables.
\end{abstract}
\mbox{}\\\\
The title of this paper is inspired by~\citet{vanBaalen2001} and in this spirit we propose as an alternative subtitle: `the epidemiology of private risk and common threat'.

\section{Introduction}\label{sec:intro}
Consider an empirical network consisting of individuals that form partnerships with other individuals. Suppose an infectious disease can be transmitted from an infectious individual to any of its susceptible partners and thus spread over the network. Consider an individual in the network at a particular point in time. We are interested in the disease status of the individual, but also in the presence of the infection in its immediate surroundings that are formed by the individual's partners. We may label this individual by listing
\begin{itemize}
\item its disease status in terms of the S, I, R classification, where, as usual, S stands for susceptible, I for infectious and R for recovered (implying immunity)
\item how many partners this individual has
\item the disease status of these partners
\end{itemize}
In this spirit, we may provide a statistical description of the network at a particular point in time by listing, for each such label, the fraction of the population carrying it.

Is it possible to predict the future spread of the disease on the basis of this statistical description? The answer is `no', simply because the precise network structure is important for transmission and we cannot recover the structure from the description. But if we are willing to make assumptions about the structure (and to consider the limit of the number of individuals going to infinity), the answer might be `yes'. And even if the true answer is still `no', we may indulge in wishful thinking and answer `to good approximation'.

When considering an outbreak of a rapidly spreading disease, we can consider the network as static. If we are willing to assume that the network is constructed by the configuration procedure~\citep{Durrett2006, vdHofstad2015}, the answer is indeed `yes'~\citep{Decreusefond2012, Barbour2013, Janson2014}. But if the disease spreads at the time scale of formation and dissolution of partnerships, we need to take these partnership dynamics into account and next indeed rely on wishful thinking (though the answer may very well be `yes'). In case of HIV, the disease spreads on the time scale of demographic turnover and this motivated our earlier work~\citep{Leung2012, Leung2015a} that also takes birth and death into account (here we know that the answer is `no', see~\citet[Appendix B]{Leung2015a}). 

In the rest of this introduction we first discuss the model formulation used and the relation between our work and existing literature. Next, we consider three different settings based on the time scales of disease spread, partnership dynamics, and demographic turnover. Individuals are decomposed into conditionally independent components (the `binding sites') and we discuss how the dynamics of these binding sites can be specified. We end the introduction with an outline of the structure of the rest of the paper.

\subsection*{Physiologically Structured Population Models}
As in our earlier paper~\citep{Leung2015a}, our model formulation is in the tradition of physiologically structured population models (PSPM~\citep{Metz1986,Diekmann1998b,Diekmann2001}). This means that we start from the notion of state at the individual level, called i-state (where i stands for individual). Model specification involves, first of all, a description of changes in time of the i-state as influenced by i-state itself and the relevant environmental variables that capture the influence of the outside world. Next the model specifies the impact of individuals on the environmental variables. Thus the feedback loop that creates density dependence, i.e.\ dependence among individuals, is described in a two step procedure. To lift the i-level model to the population level (p-level) is just a matter of bookkeeping, see~\citet{Diekmann2010} for a recent account.

In the setting considered here, i-state ranges over a finite set. As a consequence, the p-level equations are ordinary differential equations (ODE). These ODE describe, apart from death and birth of individuals, the dynamical changes of i-state, i.e.\ how individuals jump back and forth between the various states. In the spirit of the theory of Markov chains~\citep{Taylor1998}, we describe an individual not by its actual state but by the probability distribution, i.e.\ the probability of being in the various states. Equating a p-level fraction to an i-level probability provides the link between the two levels.

The approach of both earlier work and this paper is to \emph{pretend} that the label can be considered as the i-state, the information about the individual that is relevant for predicting its future. The i-state contains information about partners, but not about partners of partners. Implicitly this entails that we use a mean field description of partners of partners. We call this the `mean field at distance one' assumption. The description of partners of partners is incorporated in an environmental variable, the information about the `outside world' that is relevant for a prediction of the future of the individual. 

\begin{figure}[H]
\centering
\includegraphics[scale=0.5]{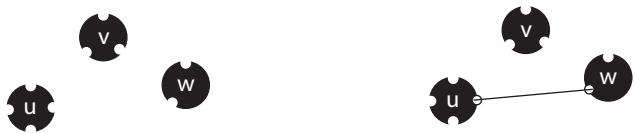}
\caption{An illustration of binding sites with three individuals $u$, $v$, and $w$. In this example, $u$, $v$, and $w$ have four, three, and two binding sites for partners, respectively. On the left, all binding sites are free. On the right, a partnership between $u$ and $w$ is formed and they both have one occupied binding site.}
\label{fig:bindingsites}
\end{figure}

A rather special feature of the models considered here is that i-state involves a number of conditionally independent components: the binding sites. An individual has binding sites for partners. Two free binding sites can be joined together to form a partnership between two individuals (see Fig.~\ref{fig:bindingsites} for an illustration). In graph theory the words `half-edge' or `stub' are often used. We think that for static networks these terms capture the essence much better than the word `binding site'. But the latter provides, in our opinion, a better description for dynamic networks. The fact that our research started with dynamic networks is responsible for our choice of terminology. 

It is attractive to model the dynamics of one binding site and next use combinatorics to describe the full i-state. It is precisely this aspect that we did not yet elaborate in~\citet{Leung2015a} but highlight now. It is precisely this aspect that uncovers the link/relationship between the work of~\citet{Lindquist2011,Leung2015a} on the one hand and the edge-based modelling approach of Volz, Miller and co-workers~\citep{Volz2007,Volz2008,Volz2009,Miller2011, Miller2013} on the other hand. 

Volz and Miller focus on the binding site (=half-edge/stub) and individual level and draw p-level conclusions by a clever use of probabilistic arguments to determine the relevant environmental variables. \citet{Lindquist2011} systematically formulate and analyse the p-level equations. In~\citet{Leung2015a} we too emphasized the p-level equations, but used the i-level version to derive an expression for $R_0$. The link between the two was established by somewhat contrived linear algebra arguments. In the present paper we build our way upwards from binding site - via individual - to population level. One of the secondary aims of this paper is to show that the systematic methodology of PSPM is also very useful when i-state space is discrete, rather than a continuum, and when i-state involves multiple identical components.

\subsection*{Three network cases}
Now, consider a network. An epidemic starts when, at some point in time, a small fraction of the population is infected from outside. Our idealized description shifts the `point in time' towards $-\infty$ while letting the fraction become smaller and smaller. In other words, our story starts `far back' in time when all individuals are still susceptible (see Appendix~\ref{initial} for elucidation). We consider three different situations, characterized by the relation between the time scales of, respectively, transmission, partnership dynamics and demographic turnover:
\begin{itemize}
\item[I] The disease dynamics are fast relative to any partnership- or demographic changes. The network is static and everyone is susceptible at time $t=-\infty$.
\item[II] The disease dynamics are on the same time scale as the partnership dynamics, but fast relative to demographic turnover. In this network individuals can acquire and lose partners over time. Everyone is susceptible at time $t=-\infty$.
\item[III] The disease dynamics and partnership- and demographic changes are on the same time scale. Here the age of an individual matters and we assume that, at birth, an individual enters the population as a susceptible without any partners. 
\end{itemize}

We assume that infection is transmitted from an infectious individual to a susceptible partner at rate $\beta$ and infectious individuals recover at rate $\gamma$ (but see Section~\ref{app:RE} for a far more general setting). We also assume that infection does \emph{not} influence the partnership dynamics or the probability per unit of time of dying in any way.

Each individual in the population is assumed to have a so-called partnership capacity $n$ which denotes the number of binding sites it has (so $n$ is the maximum number of simultaneous partners it may have). Throughout the life of the individual this partnership capacity does not change. An individual with partnership capacity $n$ can be thought of as having $n$ binding sites for partners (in Fig.~\ref{fig:bindingsites}, individuals $u$, $v$, and $w$ have partnership capacities 4, 3, and 2, respectively). We call the individual to which a binding site belongs its owner. For the purpose of this paper, we will assume that all individuals have the same partnership capacity $n$. One can easily generalize this by allowing individuals to have different partnership capacities; in that case, one only needs to average over $n$ in the correct way (see Section~\ref{app:RE} for the static case).

\subsection*{Binding sites}
An individual with partnership capacity $n$ is to some extent just a collection of $n$ binding sites. These $n$ binding sites are coupled through the disease status (or death) of their owner. We assume that this is the only manner in which the binding sites of an individual are coupled. As long as the disease status of the owner does not change (and the owner does not die), binding sites behave independently of one another and the `rules' for changes in binding site states are the same for each binding site. Obviously the latter depends on the network dynamics under consideration (either case I, II, or III). As a port to the world, a binding site can be in one of four states:
\begin{itemize}
\item 0 - free
\item 1 - occupied by a susceptible partner
\item 2 - occupied by an infectious partner
\item 3 - occupied by a recovered partner.
\end{itemize}
Here (and in the remainder of this introduction) our formulation is precise for case II while sometimes requiring minor adaptations to capture cases I and III. 

A key component of the model is the description of the dynamics of a binding site. The state of an individual is specified by listing its disease status and the states of each of its $n$ binding sites. So it makes sense to first consider a binding site as a separate and independent entity and to only take the dependence (by way of a change in the disease status of the owner) into account when we combine $n$ binding sites into one individual.

The case of a susceptible binding site (i.e.\ a binding site with a susceptible owner) is, as will become clear, far more important than the other cases. This is partly due to our assumption that all individuals start out susceptible, i.e.\ are susceptible at time $t=-\infty$ (I and II) or at birth (III). The dynamics of a susceptible binding site are described by a differential equation for the variable $x(t)=(x_i(t))$, $i=0,1,2,3$. Here $x_i$ can be interpreted as the probability that a binding site is susceptible and has state $i$ at time $t$, given that its owner does not become infected through one of its other $n-1$ binding sites (in other words, by conditioning on the individual not getting infected through its $n-1$ other binding sites, the only way the individual could get infected is through the binding site under consideration). In particular, \emph{given that its owner does not become infected through one of its other binding sites}, 
\begin{equation}\label{def:xbar}
\bar x(t)=x_0(t)+x_1(t)+x_2(t)+x_3(t)
\end{equation}
is the probability that the binding site is susceptible at time $t$ (or, in other words, that the owner is not infected along this binding site before time $t$). Accordingly, the probability that an \emph{individual} is susceptible at time $t$ is equal to
\begin{equation}\label{def:SI}
\bar x(t)^n.
\end{equation}

In order to arrive at a closed system of equations for $x$, we need to go through several steps. The variable $x$ contains information about a partner. Consequently the dynamics of $x$ is partly determined by partners of partners, hence by one or more environmental variables. The `mean field at distance one' assumption yields expressions for environmental variables in terms of subpopulation sizes (for a given label, the corresponding subpopulation size is the fraction of the population that carries this label). In turn, p-level fractions can be expressed in terms of i-level probabilities. And since a susceptible individual is in essence a collection of $n$ conditionally i.i.d.\ binding sites, we can use combinatorics to express i-level probabilities in terms of binding-site-level probabilities as incorporated in $x$. 

\begin{figure}[H]
\centering
\includegraphics[scale=0.2]{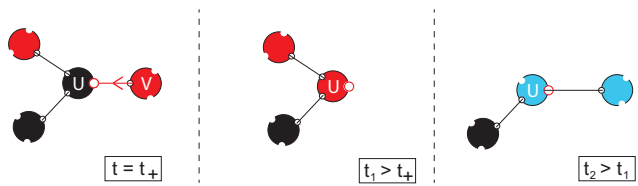}
\caption{An illustration of the exceptional binding site. Susceptible, infectious, and recovered individuals are displayed in black, red, and blue, respectively. Three time points in the life of individual $u$ are displayed. Suppose $u$ is susceptible and becomes infected by an infectious partner $v$ at time $t_+$. From that moment on, the binding site along which transmission occurred is the exceptional binding site. This binding site remains exceptional throughout $u$'s life and no other binding site can become exceptional, regardless of whether or not $v$ is still a partner or $u$ is still infectious.}
\label{fig:exceptional}
\end{figure}

The exchangeability of the binding sites is broken by the infection event. There is exactly one binding site along which infection took place, viz.\ the binding site occupied by the individual's epidemiological parent, and for this binding site we know with certainty that it is in state 2 at time of infection $t_+$. We call the binding site through which the change in the owner's disease status occurred the `exceptional' binding site. The other $n-1$ binding sites are i.i.d.\ and, at time $t_+$, they are distributed according to $x(t_+)$. Recovery (and death) is an event that occurs at a constant rate for an infectious individual so independent of binding site states. Therefore, also after recovery, there remains exactly one exceptional binding site, viz.\ the one through which transmission occurred. See also Fig.~\ref{fig:exceptional} for an illustration of the exceptional binding site. 

\subsection*{Structure of the paper}
In Sections~\ref{sec:static},~\ref{sec:intermediate}, and~\ref{sec:demography} below, we will discuss the three network model cases I, II, and III separately. For each of the three cases we will explain how the model can be formulated and described in terms of susceptible binding sites. By considering the susceptible binding site perspective we can write a closed system of only a few equations that fully determine the dynamics of i-level probabilities and p-level fractions. This system is then used to determine epidemiological quantities of interest: $R_0$, $r$, the final size (in cases I and II), and the endemic steady state (in case III). In all three cases, an explicit expression can be given for $R_0$. In case I, one can derive a simple scalar equation for the final size. In cases II and III, we could only  implicitly characterize the final size and endemic equilibrium, respectively.

In Section~\ref{sec:static} case I of a static network is considered. This is the simplest case among the three. The relative simplicity allows for the derivation of an ODE system for susceptible binding sites directly from the interpretation. This will be the first way in which we formulate the model for this case. But case I will also serve to illustrate the systematic procedure for model formulation in the spirit of PSPM. This systematic procedure allows us to connect the three different levels, viz.\ (1) binding sites, (2) individuals, and (3) the population, to each other. 

In network case I, since it is relatively simple, one can derive a one-dimensional renewal equation from which $R_0$, $r$, and the final size almost immediately follow. This renewal equation will be treated in Section~\ref{app:RE} for a much more general class of infectious disease models than only SIR.

Part of the systematic procedure in cases II and III focuses on infectious binding sites. We use case I to illustrate the model formulation concerning infectious (and recovered) binding sites, even though, for case I these are not needed to obtain a closed system for susceptible binding sites. However, depending on the network features of interest (e.g.\ fractions of infectious individuals) one may still want to consider infectious (and recovered) binding sites. 

In network cases II and III, there are also network dynamics in absence of infection due to partnership changes (and demographic changes). We will only describe the essential characteristics of the network dynamics that we use in this paper. Certainly, much more can be said about the networks in absence of infection~\citep{Leung2012}.  

Finally, in Section~\ref{sec:discussion} we discuss the issues that we have encountered in the three different network cases and pose some open problems. We end the discussion by considering a few generalizations that can easily be implemented using the systematic model formulation of Section~\ref{sec:steps}.

\section{Part I: static network}\label{sec:static}
\subsection{Model formulation}\label{sec:modelI}
We derive a closed system of ODE for $x$ purely on the basis of the interpretation of binding sites (without explicitly taking into account i-level probabilities or p-level fractions). The relatively simple setting of a static network allows us to do so. We are able to consider a binding site as a separate and independent entity all throughout its susceptible life. Implicitly, this uses~\eqref{eq:i2p} below. One can show that the system of ODE for $x$ indeed captures the appropriate large population limit of a stochastic SIR epidemic on a configuration network. This requires quite some work; see~\citep{Decreusefond2012,Barbour2013,Janson2014}.

Consider a susceptible binding site and assume its owner does not become infected through one of its other $n-1$ binding sites for the period under consideration. If a susceptible binding site is in state 2, it can become infected by the corresponding infectious partner. This happens at rate $\beta$ and when it happens, the binding site is no longer susceptible so it `leaves' the $x$-system. It is also possible that the infectious partner recovers. This happens at rate $\gamma$. Finally, there is the possibility that a susceptible partner of a susceptible binding site becomes infectious (corresponding to a transition from state 1 to state 2). The rate at which this occurs depends on the number of infectious partners that this susceptible partner has. So here we use the mean field at distance one assumption: we average over all possibilities at the p-level to obtain one rate at which a susceptible partner of a susceptible binding site becomes infected. More specifically, we assume that there is a rate $\beta\Lambda_-(t)$ at which a susceptible partner of a susceptible binding site becomes infected at time $t$. Here $\Lambda_-(t)$ has the interpretation of the expected number of infectious partners of a susceptible partner of a susceptible individual.

Then, putting together the various assumptions described above, the dynamics of $x$ is governed by the following system (please note that the environmental variable $\Lambda_-$ is a p-level quantity that we have yet to specify):
\begin{align}\label{eq:ODExI}
\frac{dx(t)}{dt}&=M\big(\Lambda_-(t)\big)x(t),
\end{align}
with `far past' conditions
\begin{align*}
x_1(-\infty)&=1,\quad x_2(-\infty)=0=x_3(-\infty),
\end{align*}
and
\begin{align}\label{eq:MI}
M(\Lambda_-)&=\begin{pmatrix}-\beta\Lambda_- &0&0\\\beta\Lambda_-&-(\beta+\gamma)&0\\0&\gamma&0\end{pmatrix}.
\end{align}
To express $\Lambda_-$ in terms of $x$ we use the interpretation. Consider a susceptible partner $v$ of a susceptible individual $u$. Then, since $u$ is susceptible, we know that $v$ has at most $n-1$ binding sites that are possibly in state 2 (i.e.\ occupied by infectious partners). Since $v$ is known to be susceptible, also all its binding sites are susceptible (in the sense that their owner $v$ is). The probability that a binding site is susceptible at time $t$ is $\bar x$ with 
\begin{equation}\label{def:xbarI}
\bar x(t)=x_1(t)+x_2(t)+x_3(t)
\end{equation} 
(recall~\eqref{def:xbar} and note that in case I we have $x_0(t)=0$). The probability that a binding site is in state 2, given that the binding site is susceptible, is $x_2(t)/\bar x(t)$. Therefore,
\begin{equation}\label{eq:Lambda-I}
\Lambda_-(t)=(n-1)\frac{x_2(t)}{\bar x(t)}.
\end{equation}

By inserting~\eqref{eq:Lambda-I} into~\eqref{eq:ODExI} we find that the $x$-system is fully described by an ODE system in terms of the $x$-variables only:
\begin{equation}\label{eq:ODExI_final}
\begin{aligned}
x_1'&=-\beta (n-1)\frac{x_2}{\bar x}x_1\\
x_2'&=\beta (n-1)\frac{x_2}{\bar x}x_1-(\beta+\gamma)x_2\\
x_3'&=\gamma x_2,
\end{aligned}
\end{equation}
with `far past' conditions
\begin{equation*}
x_1(-\infty)=1, \quad x_2(-\infty)=0=x_3(-\infty).
\end{equation*}

\begin{remark}
In the pioneering paper~\citep{Volz2008} an equivalent system of three coupled ODE was introduced to describe the binding-site level of the model. The variables of Volz are connected to our $x$-system as follows: $\theta=\bar x$, $p_S=x_1/\bar x$ and $p_I=x_2/\bar x$.
\end{remark}

\subsection{Systematic procedure for closing the feedback loop}\label{sec:steps}
Before analyzing~\eqref{eq:ODExI_final} in the next section, we describe a systematic procedure, consisting of five steps, for deriving the complete model formulation. A key aim is to rederive the crucial relationship~\eqref{eq:Lambda-I} in a manner that can be extended to the dynamic networks. Thus the present section serves to prepare for a quick and streamlined presentation of the cases II and III in Sections~\ref{sec:intermediate} and~\ref{sec:demography}, respectively. The various steps reveal the relation between binding site probabilities, i-level probabilities and p-level fractions. In addition we introduce some notation.

\paragraph{step 1. Susceptible binding sites: $x$-probabilities}\hfill\\
The first step is to describe the dynamics of $x$ while specifying the environmental variable $\Lambda_-$ only conceptually, i.e.\ in terms of the interpretation. We then arrive at system~\eqref{eq:ODExI}-~\eqref{eq:MI}.

\mbox{}\\
Next, we introduce $P_{(d, \B k)}(t)$, denoting the fraction of the population with label $(d, \B k)$. Here $\B{k}=(k_1, k_2, k_3)$ denotes the number of partners of an individual with each of the different disease statuses, i.e.\ $k_1$ susceptible, $k_2$ infectious, and $k_3$ recovered partners. Furthermore, $d\in\{-, +, \ast\}$ denotes the disease status of the individual itself, with $-$ corresponding to S, $+$ to I, and $\ast$ to R. 
\mbox{}\\\\
In the second step, the environmental variable $\Lambda_-$ is, on the basis of its interpretation, redefined in terms of p-level fractions $P_{(-, \B k)}(t)$.

\begin{figure}[H]
\centering
\includegraphics[scale=0.25]{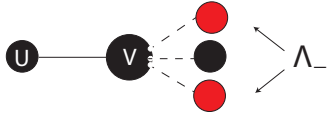}
\caption{The susceptible partner $v$ of a susceptible individual $u$ has a mean number of infectious partners $\Lambda_-$.}
\label{fig:lambda_min}
\end{figure}

\paragraph{step 2. Environmental variables: definition in terms of p-level fractions}\hfill\\
The mean field at distance one assumption concerns the environmental variable $\Lambda_-$. This variable is interpreted as the mean number of infectious partners of a susceptible individual that has at least one susceptible partner (see also Fig.~\ref{fig:lambda_min}). We define it in terms of p-level fractions as follows:
\begin{equation}\label{def:Lambda-}
\Lambda_-(t)=\sum_{\B{m}} m_2\ \frac{m_1P_{(-,\B{m})}(t)}{\sum_{\B{k}}k_1P_{(-,\B{k})}(t)}.
\end{equation}
Here the sums are over all possible configurations of $\B{m}$ and $\B{k}$ with $0\leq m_1+m_2+m_3\leq n$, $0\leq k_1+k_2+k_3\leq n$. The second factor in each term of this sum denotes the probability that a susceptible partner of a susceptible individual is in state $(-,\B{m})$. The number of infectious partners is then given by $m_2$, and we find the expected number of infectious partners $\Lambda_-$ by summing over all possibilities. 

\noindent In the third step, we let $p_{(-,\B k)}(t)$ denote the probability that an individual is in state $(-,\B k)$ at time $t$. This i-level probability can be expressed in terms of $x$-probabilities.

\paragraph{step 3. i-level probabilities in terms of $x$-probabilities}\hfill\\
We need to take into account the number of possible configurations of the individual's binding sites such that there are exactly $k_1$ binding sites in state 1, $k_2$ binding sites in state 2, (and then automatically, there are $k_3=n-k_1-k_2$ binding sites in state 3). The number of possibilities is equal to 
\begin{equation*}\label{eq:binomI}
\binom{n}{k_1+k_2}\binom{k_1+k_2}{k_1}=\frac{n!}{k_1!\, k_2!\,k_3!},
\end{equation*}
The probability to have a specific configuration of the $n$ binding sites in the different states is obtained by simply multiplying the $x$-probabilities:
\begin{equation*}
x_1^{k_1}x_2^{k_2}x_3^{k_3}.
\end{equation*}
Therefore, 
\begin{equation}\label{def:pxI}
p_{(-,\B{k})}(t)=\frac{n!}{k_1!\, k_2!\,k_3!}\left(x_1^{k_1}\, x_2^{k_2}\, x_3^{k_3}\right)(t)
\end{equation} 
is the probability that an individual is, at time $t$, susceptible with $k_1$ susceptible, $k_2$ infectious, and $k_3$ recovered partners. The solution of the $x$-system then gives us a complete Markovian description of the i-state dynamics of susceptible individuals.

\mbox{}\\
In this setting of a static network age does not play a role. Therefore, i-level probabilities can immediately be linked to p-level fractions in step 4 below.

\paragraph{step 4. p-level fractions in terms of i-level probabilities}\hfill\\
The i-level probabilities and p-level fractions coincide, i.e.\
\begin{equation}\label{eq:i2p}
P_{(d,\B k)}(t)=p_{(d,\B k)}(t),
\end{equation}
$d\in\{-,+,\ast\}$. In a way, individuals are interchangeable as they all start off in the same state at $t=-\infty$.

\mbox{}\\
Finally in the last step, by combining steps 2, 3, and 4, we can express $\Lambda_-$ in terms of the $x$-probabilities.

\paragraph{step 5. Environmental variables in terms of $x$-probabilities (combining 2, 3, 4)}\hfill\\
By combining~\eqref{eq:i2p}, and~\eqref{def:pxI} we find that
$\sum_{\B{m}} m_2m_1P_{(-,\B{m})}(t)=n(n-1)\left(x_1x_2\bar x^{n-2}\right)(t)$ and $\sum_{\B{k}}k_1P_{(-,\B{k})}(t)=n \left(x_1\bar x^{n-1}\right)(t)$. Then definition~\eqref{def:Lambda-} yields the same expression for $\Lambda_-$ as~\eqref{eq:Lambda-I}.

\mbox{}\\
Finally, steps 1 to 5 together yield the closed system~\eqref{eq:ODExI_final} of ODE for $x$. The dynamics of the $1/2(n+1)(n+2)$ i-level probabilities $p_{(-,\B k)}(t)$ are fully determined by the system of three ODE for $x$. We can use this three-dimensional system of ODE to determine $r$, $R_0$, and the final size as we will show in Section~\ref{sec:R0I}. In this particular case of a static network, we can do even better by considering one renewal equation for $\bar x$. This one equation then allows us to determine the epidemiological quantities as well. This is the topic of Section~\ref{app:RE} where we consider epidemic spread on a static configuration network in greater generality.

\begin{remark}\label{rk:ODEpI}
One obtains the p-level ODE system by differentiation of~\eqref{def:pxI} and use of~\eqref{eq:ODExI_final} and~\eqref{eq:i2p}. In doing so, one obtains a system of $1/2(n+1)(n+2)$ ODE for the p-level fractions concerning individuals with a $-$ disease status:
\begin{align*}
\frac{dP_{(-, k_1,k_2, k_3)}}{dt}&=-(\beta k_2+\gamma k_2+\beta\Lambda_-k_1)P_{(-,k_1,k_2, k_3)}+\gamma P_{(-,k_1,k_2+1, k_3-1)}\nonumber\\
&\phantom{=\ }+\beta\Lambda_-(k_1+1)P_{(-,k_1+1,k_2-1, k_3)},
\end{align*}
$k_1+k_2+k_3=n$, with $\Lambda_-$ defined by~\eqref{def:Lambda-} (compare with~\citet[eq. (13)]{Lindquist2011}).
\end{remark}

\subsection{The beginning and end of an epidemic: $R_0$, $r$, and final size}\label{sec:R0I}\label{sec:finalsize_I}
In this section we consider the beginning and end of an epidemic. We first focus on $R_0$ and $r$, so on the start of an epidemic.

Note that we can very easily find an expression for $R_0$ from the interpretation: when infected individuals are rare, a newly infected individual has exactly $n-1$ susceptible partners. It infects one such partner before recovering from infection with probability $\beta/(\beta+\gamma)$. Therefore, the expected number of secondary infections caused by one newly infected individual is
\begin{equation}\label{eq:R0I}
R_0=\frac{\beta(n-1)}{\beta+\gamma}.
\end{equation}
However, even though there should be no doubt about it, this does not yield a \emph{proof} that this expression is indeed a threshold parameter with threshold value one for the stability of the disease free steady state of the p-level system. In order to provide a proof \emph{and} to prepare for cases II and III, we now derive $R_0$ and $r$ from the binding site system~\eqref{eq:ODExI_final}. 

Note that the p-level fractions $P_{(-,\B k)}(t)$ can be fully expressed in terms of the binding site level probabilities $x_i$ (eqs.~\eqref{eq:i2p} and~\eqref{def:pxI}). Furthermore, the $P_{(-,\B k)}(t)$ fractions, i.e.\ the fractions concerning individuals with a $-$ disease status, form a closed system. Therefore, a threshold parameter for the disease free steady state of the binding-site system $x$ is also a threshold parameter for the disease free steady state of the p-level system. (This argument extends to the dynamic network cases II and III in Sections~\ref{sec:intermediate} and~\ref{sec:demography})

Linearization of system~\eqref{eq:ODExI_final} in the disease free steady state $\tilde x_1=1$, $\tilde x_2=0=\tilde x_3$, yields a decoupled ODE for the linearization of the ODE for $x_2$. To avoid any confusion, let $\hat x_2$ denote the \emph{linearized} $x_2$ variable. Then the linearization yields
\begin{equation*}
\hat x_2'=\beta(n-1)\hat x_2-(\beta+\gamma)\hat x_2,
\end{equation*}
with `far past' condition $\hat x_2(-\infty)=0$. In particular, the right-hand side of the ODE for $\hat x_2$ depends only on $\hat x_2$. 

To illustrate the method used in case II and III in Sections~\ref{sec:R0II} and~\ref{sec:R0III}, we derive expressions for $R_0$ and $r$ from a special form of the characteristic equation. Variation of constants for the ODE of $\hat x_2$ yields
\begin{equation*}
\hat x_2(t)=\int_0^\infty e^{-(\beta+\gamma)\tau}\beta(n-1)\hat x_2(t-\tau)d\tau.
\end{equation*}
Substituting the ansatz $\hat x_2(t)=e^{\lambda t}$ yields the characteristic equation
\begin{equation*}
1=\int_0^\infty \beta e^{-(\beta+\gamma)\tau}(n-1)e^{-\lambda \tau}d\tau.
\end{equation*}
Then there is a unique real root to this equation for $\lambda$ that we denote by $r$ and call the Malthusian parameter. Evaluating the integral we find that $r=\beta(n-1)-(\beta+\gamma)$. Likewise, we can derive the expression~\eqref{eq:R0I} for $R_0$ by evaluating the integral with $\lambda=0$. 

Next, we consider the final size. We do so by considering the dynamics of $\bar x$ defined in~\eqref{def:xbarI}. Recall~\eqref{def:SI}, i.e.\ the probability that an individual is susceptible at time $t$, is given by $\bar x(t)^n$. We observe that, by~\eqref{eq:i2p}, $\bar x(t)^n$ is also equal to the fraction of susceptible individuals in the population at time $t$. (Alternatively, one can show that $\sum_{\B k}P_{(-,\B k)}(t)=\bar x(t)^n$ by combining~\eqref{eq:i2p} and \eqref{def:pxI}.) In fact, it is possible to describe the dynamics of $\bar x$ in terms of only $\bar x$ itself. This was first observed in~\citet{Miller2011b}, where the Volz equations of~\citep{Volz2008} were taken as a starting point. The most important observation is the consistency relation 
\begin{equation}\label{eq:x1barx}
x_1=\bar x^{n-1}.
\end{equation}  
We can use the interpretation to derive~\eqref{eq:x1barx}; $x_1$ is the probability that a susceptible binding site with owner $u$ is occupied by a susceptible partner $v$, $\bar x^{n-1}$ is the probability that $v$ is susceptible given that it is a partner of a susceptible individual $u$ (see also~\eqref{eq:x1consistency} below). 

Then, using~\eqref{eq:x1barx} together with algebraic manipulation of the ODE system~\eqref{eq:ODExI_final} (see~\citep{Miller2011b} for details), one is able to find a decoupled equation for $\bar x$:
\begin{equation}\label{eq:barx}
\bar x'=\beta\bar x^{n-1}-(\beta+\gamma)\bar x+\gamma.
\end{equation}
The fraction of susceptible individuals at the end of the outbreak is determined by the probability $\bar x(\infty)$. Since $\bar x$ satisfies~\eqref{eq:barx} and $\bar x(\infty)$ is a constant, we find that necessarily $\bar x(\infty)$ is the unique solution in $(0,1)$ of
\begin{equation}\label{eq:app_finalsize}
0=\beta\bar x(\infty)^{n-1}-(\beta+\gamma)\bar x(\infty)+\gamma
\end{equation}
if $R_0>1$. The final size is given by
\begin{equation*}
1-\bar x(\infty)^n.
\end{equation*}

In Section~\ref{app:RE} we show that one can actually describe the dynamics of the probability $\bar x$ for deterministic epidemics on configuration networks for a much larger class of submodels for infectiousness. The SIR infection that we consider here is a very special case of the situation considered in Section~\ref{app:RE}. There we show that it is possible to derive a renewal equation for $\bar x$. The final size equation is then obtained by simply taking the limit $t\to\infty$. We highly recommend reading Section~\ref{app:RE} to understand the derivation of the renewal equation for $\bar x$ based on the interpretation of the model (with a minimum of calculations).

\subsection{After susceptibility is lost}\label{sec:infI}
In the preceding section we have seen that the $x$-system~\eqref{eq:ODExI_final} for susceptible binding sites is all that is needed to determine several epidemiological quantities of immediate interest. On the other hand, we might not only be interested in the fraction~\eqref{def:SI} of susceptibles in the population, but also in the dynamics of i-level probabilities $p_{(d,\B k)}(t)$ (and likewise p-level fractions $P_{(d,\B k)}(t)$ given by~\eqref{eq:i2p}) for $d=+,\ast$.

So what happens after an individual becomes infected? We work out the details for infectious individuals and only briefly describe recovered individuals. Again, we are able to formulate the model following steps 1-5 of Section~\ref{sec:steps} (where the word `susceptible' should be replaced by `infectious' or `recovered' whenever appropriate and step 3 should be replaced by a slightly different step 3', but we will come back to this later on in this section). But now we need to take into account the exceptional binding site, i.e.\ the binding site through which infection was transmitted to the owner (see also Fig.~\ref{fig:exceptional}). 

In \emph{step 1} one considers the dynamics of infectious binding sites, i.e.\ binding sites having infectious owners. Suppose that the owner became infected at time $t_+$ and that it does not recover in the period under consideration. Let $y_i^\text{e}(t\mid t_+)$ denote the probability for the exceptional binding site to be in state $i$ at time $t$, $i=1,2,3$. Similarly, $y_i(t\mid t_+)$ denotes the probability for a non-exceptional binding site to be in state $i$ at time $t$, $i=1,2,3$. Here the probabilities are defined only for $t\geq t_+$. Note that $y$ and $y^\text{e}$ are probability vectors, i.e.\ the components are nonnegative and sum to one. 

Instead of `far past' conditions we now have to take into account the distribution of binding site states at time of infection $t_+$. Whether or not an infectious binding site is exceptional has an influence on the state it has at epidemiological birth. Indeed, the exceptional binding site is in state $2$ at time $t_+$ with probability 1, while the distribution of the state of a non-exceptional binding site at time $t_+$ is given by $x(t_+)/\bar x(t_+)$, i.e.\ we have boundary conditions
\begin{equation}\label{static:BC}
\begin{aligned}
y_1^\text{e}(t_+\mid t_+)&=0, &y_1(t_+\mid t_+)&=x_1(t_+)/\bar x(t_+),\\
y^\text{e}_2(t_+\mid t_+)&=1, &y_2(t_+\mid t_+)&=x_2(t_+)/\bar x(t_+),\\
y^\text{e}_3(t_+\mid t_+)&=0, &y_3(t_+\mid t_+)&=x_3(t_+)/\bar x(t_+).
\end{aligned}
\end{equation}

\begin{figure}[H]
\centering
\includegraphics[scale=0.25]{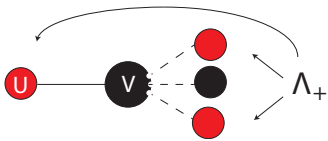}
\caption{The susceptible partner $v$ of an infectious individual $u$ has a mean number of infectious partners $\Lambda_+$ (note that this number is always larger or equal to 1 since $u$ is a partner).}
\label{fig:lambda_plus}
\end{figure}

The mean field at distance one assumption again plays a role. Here, we need to deal with the environmental variable $\Lambda_+$ that is defined as the expected number of infectious partners of a susceptible partner of an infectious individual (see also Fig.~\ref{fig:lambda_plus} and compare with Fig.~\ref{fig:lambda_min}). We can redefine $\Lambda_+$ in terms of p-level fractions $P_{(-,\B k)}$ for susceptible individuals:
\begin{equation}\label{def:Lambda+}
\Lambda_+(t)=\sum_{\B m} m_2\frac{m_2P_{(-, \B m)}(t)}{\sum_{\B k} k_2P_{(-,\B k)}(t)}.
\end{equation} 
In particular, once again, $\Lambda_+$ can be expressed in terms of $x$ by combining steps 2, 3, and 4. Using~\eqref{def:Lambda+},~\eqref{eq:i2p}, and~\eqref{def:pxI} we find that
\begin{equation}\label{eq:lambda+2x}
\Lambda_+(t)=1+(n-1)\frac{x_2(t)}{\bar x(t)}
\end{equation}
(alternatively, one can find the same expression for $\Lambda_+$ in terms of $x$-probabilities directly from the interpretation, exactly as before in the case of $\Lambda_-$).

The rates at which changes in the states (1, 2, 3) of infectious binding sites occur is the same for each binding site, including the exceptional one. There is a rate $\gamma$ at which an infectious partner of an infectious binding site recovers (this corresponds to a change in state from 2 to 3). And there is a rate at which a susceptible partner of an infectious binding site becomes infected (either along the binding site under consideration or by one of its other infectious partners) corresponding to a change in state from 1 to 2. The rate at which this occurs is $\beta\Lambda_+$ where $\Lambda_+$ is defined by~\eqref{def:Lambda+} and hence~\eqref{eq:lambda+2x}.

Recall that we condition on the infectious binding site under consideration not recovering, therefore, these are all state changes that can occur. So we find that the dynamics of $y$ and $y^e$ are described by the same ODE system
\begin{equation}\label{eq:ODEyI}
\begin{aligned}
%\frac{d y^\text{e}(t\mid t_+)}{d t}&=M_+\big(\Lambda_+(t)\big)y^\text{e}(t\mid t_+),\\
\frac{d y(t\mid t_+)}{d t}&=M_+\big(\Lambda_+(t)\big)y(t\mid t_+),
\end{aligned}
\end{equation}
with 
\begin{align*}
M_+(\Lambda_+)&=\begin{pmatrix}-\beta\Lambda_+ &0&0\\\beta\Lambda_+&-\gamma&0\\0&\gamma&0\end{pmatrix},
\end{align*}
and case specific boundary conditions~\eqref{static:BC}. Observe that this means that $y^\text{e}_1(t\mid t_+)=0$ for all $t\geq t_+$.This also immediately follows from the interpretation: at time $t_+$, the binding site is occupied by an infectious partner, the network is static, and an infectious individual can not become susceptible again. 
\mbox{}\\\\
Next, we turn to \emph{infectious} individuals. Compared to susceptible i-level probabilities, it is more involved to express infectious i-level probabilities in terms of $y^\text{e}$- and $y$-probabilities. Therefore, we first consider conditional i-level probabilities before finding an expression for the unconditional probabilities. We replace step 3 by step 3'.

\paragraph{step 3' Infectious i-level probabilities in terms of $y$ and $y^\text{e}$}\hfill\\
We let $\phi_{(+,\B k)}(t\mid t_+)$ denote the probability that an infectious individual, infected at time $t_+$, is in state $(+,\B k)$ at time $t$, given no recovery. As in the case of a susceptible individual, we count the number of different configurations for the $n$ binding sites of the individual (and we find the same expression as in the case of a susceptible individual). Next, we need to take into account that there is one exceptional binding site, and the probability that it is in state $i$ is $y_i^\text{e}$ (note that this is equal to zero for $i=1$). The other $n-1$ binding sites are i.i.d. Suppose the exceptional binding site is in state 2, then the number of possible configurations of the individual's $n-1$ non-exceptional binding sites such that there are exactly $k_1$ in state 1, $k_2-1$ in state 2, and $k_3$ in state 3 is 
\begin{equation*}
\binom{n-1}{k_1+k_2-1}\binom{k_1+k_2-1}{k_1}=\frac{(n-1)!}{k_1!\, (k_2-1)!\, k_3!}.
\end{equation*}
The probability to have a specific configuration of the $n-1$ binding sites in the different states is obtained by multiplying the $y$-probabilities:
\begin{equation*}
y_1^{k_1}y_2^{k_2-1}y_3^{k_3}.
\end{equation*}
We can do the same when the exceptional binding site is in state 3. Taking into account both possible states (2 and 3) for the exceptional binding site, we obtain
\begin{align}\label{eq:infstatic1}
\phi_{(+,\B k)}(t\mid t_+)=\frac{n!}{k_1!\, k_2!\, k_3!}&\left(\frac{k_2}n\,y_2^\text{e}\ y_1^{k_1}y_2^{k_2-1}y_3^{k_3}+\frac{k_3}n\,y_3^\text{e}\ y_1^{k_1}y_2^{k_2}y_3^{k_3-1}\right)(t\mid t_+).
\end{align}
Note that $\phi_{(+,\B k)}(t\mid t_+)=0$ for $\B k=(n, 0,0)$, i.e.\ for all $t\geq t_+$ at least one partner is not susceptible. 

A susceptible individual becomes infected at time $t_+$ if infection is transmitted to this individual through one of its $n$ binding sites. Infection is transmitted at rate $\beta$. Therefore, the force of infection at time $t_+$, i.e.\ the rate at which a susceptible individual becomes infected at time $t_+$, equals $\beta n \frac{x_2}{\bar x}(t_+)$ and consequently the incidence at time $t_+$, i.e.\ the fraction of the population that becomes, per unit of time, infected at time $t_+$, equals 
\begin{equation}\label{eq:incidence}
\beta n\left(\frac{x_2}{\bar x}\,\bar x^{n}\right)(t_+)=\beta nx_2\bar x^{n-1}(t_+)
\end{equation}
(recall that $\bar x^n$ is the fraction of the population that is susceptible).

Furthermore, an infectious individual that is infected at time $t_+$ is still infectious at time $t$ if it does not recover in the period $(t_+, t)$. Since the infectious period of an individual is assumed to be exponentially distributed with rate $\gamma$, the probability that this happens is 
\begin{equation}\label{eq:no_rec}
e^{-\gamma(t-t_+)}. 
\end{equation}

We then find an expression for the unconditional i-level probabilities $p_{(+,\B k)}(t)$ that a randomly chosen individual is in state $(+,\B k)$  at time $t$ in terms of infectious binding site probabilities and the history of susceptible binding site probabilities:
\begin{equation}\label{def:inf_i}
p_{(+,\B k)}(t)=\int_{-\infty}^te^{-\gamma(t-t_+)}\beta nx_2\bar x^{n-1}(t_+)\phi_{(+,\B k)}(t\mid t_+)dt_+,
\end{equation}
where $\phi_{(+,\B k)}(t\mid t_+)$ is given by~\eqref{eq:infstatic1}. The i-level probabilities $p_{(+,\B k)}(t)$ are lifted to the p-level by~\eqref{eq:i2p}.

\mbox{}\\In this way we can use infectious binding sites as building blocks for infectious individuals. We see that $y$ and $y^\text{e}$ explicitly depend on the dynamics of $x$ through the boundary conditions~\eqref{static:BC} and the environmental variable $\Lambda_+$~\eqref{eq:lambda+2x}. In addition, $x_2$ plays a role in determining the time of infection of an individual. 

\begin{remark}\label{rk:ODEp+I}
Similar to the ODE system for $-$ individuals considered in Remark~\ref{rk:ODEpI}, one obtains the p-level ODE system by differentiation of~\eqref{def:inf_i} and use of~\eqref{eq:ODEyI},~\eqref{def:pxI} and~\eqref{eq:i2p}. In doing so, one obtains a system of $1/2(n+1)(n+2)$ ODE for the p-level fractions concerning individuals with a $+$ disease status:
\begin{align*}
\frac{dP_{(+, k_1,k_2, k_3)}}{dt}&=\beta k_2 P_{(-,k_1,k_2,k_3)}-(\gamma k_2+\gamma+\beta\Lambda_+k_1)P_{(+,k_1,k_2, k_3)}+\gamma P_{(+,k_1,k_2+1, k_3-1)}\nonumber\\
&\phantom{=\ }+\beta\Lambda_+(k_1+1)P_{(+,k_1+1,k_2-1, k_3)},
\end{align*}
$k_1+k_2+k_3=n$, with $\Lambda_+$ defined by~\eqref{def:Lambda-} (compare with~\citet[eq.\ (13)]{Lindquist2011}).
\end{remark}

In case of recovered individuals, one considers their binding sites and first conditions on time of infection $t_+$ and time of recovery $t_\ast$. Again one needs to distinguish between the exceptional and the non-exceptional binding sites. The dynamics of recovered binding sites are described by taking into account the mean field at distance one assumption for the mean number $\Lambda_\ast$ of infectious partners of a susceptible partner of a recovered individual. Boundary conditions are given by the $y(t_\ast\mid t_+)$ and $y^\text{e}(t_\ast\mid t_+)$ for non-exceptional and exceptional binding sites, i.e.
\begin{alignat*}{2}
z_1^\text{e}(t_\ast\mid t_+, t_\ast)&=0,\qquad &&z_1(t_\ast\mid t_+, t_\ast)=y_1(t_\ast\mid t_+),\\
z^\text{e}_2(t_\ast\mid t_+, t_\ast)&=y_2^\text{e}(t_\ast\mid t_+),\qquad &&z_2(t_\ast\mid t_+, t_\ast)=y_2(t_\ast\mid t_+),\\
z^\text{e}_3(t_\ast\mid t_+, t_\ast)&=y_3^\text{e}(t_\ast\mid t_+),\qquad &&z_3(t_\ast\mid t_+, t_\ast)=y_3(t_\ast\mid t_+).
\end{alignat*}
The dynamics for $z$ and $z^\text{e}$ can be described by a system of ODE identical to the ODE systems for $y$ and $y^\text{e}$, but with $\Lambda_+$ replaced by $\Lambda_\ast$. The environmental variable $\Lambda_\ast$ is given by
\begin{equation}\label{def:Lambda*}
\Lambda_\ast(t)=\sum_{\B m} m_2\frac{m_3P_{(-, \B m)}(t)}{\sum_{\B k} k_3P_{(-,\B k)}(t)}.
\end{equation}
 By combining~\eqref{def:Lambda*} with~\eqref{eq:i2p} and~\eqref{def:pxI} we find
\begin{equation}\label{def:Lambda*x}
\Lambda_\ast(t)=(n-1)\frac{x_3(t)}{\bar x(t)}.
\end{equation}
We find an expression for the probability $\psi_{(\ast,\B k)}(t\mid t_+, t_\ast)$ that a recovered individual, infected at time $t_+$ and recovered at time $t_\ast$, is in state $(\ast, \B k)$ at time $t\geq t_\ast$, in terms of $z$ and $z^\text{e}$ probabilities for recovered binding sites with the same reasoning as for $\phi_{(+,\B k)}(t\mid t_+)$ (one can simply replace $\phi$ by $\psi$, $y_i$ by $z_i$, and $y_i^\text{e}$ by $z_i^\text{e}$ in~\eqref{eq:infstatic1}). Then, to arrive at an expression for the unconditional probability $p_{(\ast,\B k)}(t)$, we again need to take into account the incidence $\beta nx_2\bar x^{n-1}(t_+)$ at $t_+$. The probability that recovery does not occur in the time interval $(t_+, t_\ast)$ is given by $e^{-\gamma(t_\ast-t_+)}$ and the rate at which an infectious individual recovers is $\gamma$, therefore
\begin{equation}\label{def:staticPrec}
P_{(\ast,\B k)}(t)=p_{(\ast,\B k)}(t)=\int_{-\infty}^t\int_{-\infty}^{t_\ast}\gamma e^{-\gamma(t_\ast-t_+)}\beta nx_2\bar x^{n-1}(t_+)\psi_{(\ast,\B k)}(t\mid t_+, t_\ast)dt_+dt_\ast,
\end{equation}
where the first equality in~\eqref{def:staticPrec} follows from~\eqref{eq:i2p}.

\subsection{The renewal equation for the Volz variable}\label{app:RE}
So far we dealt with the SIR situation, where an individual becomes infectious immediately upon becoming infected and stays infectious for an exponentially distributed amount of time, with rate parameter $\gamma$, hence mean $\gamma^{-1}$. During the infectious period any susceptible partner is infected with rate (=probability per unit of time) $\beta$. 

Here we incorporate randomness in infectiousness via a variable $\xi$ taking values in a set $\Omega$ according to a distribution specified by a measure $m$ on $\Omega$. This sounds abstract at first, but hopefully less so if we mention that the SIR situation corresponds to
\begin{align*}
\Omega&=(0,\infty),\\
m(d\xi)&=\gamma e^{-\gamma\xi}d\xi,
\end{align*}
with $\xi$ corresponding to the length of the infectious period. In this section we only consider the setting where the `R' characteristic holds, i.e.\ after becoming infected, individuals can not become susceptible for infection any more.

In order to describe how the probability of transmission to a susceptible partner depends on $\xi$, we need the auxiliary variable $\tau$ corresponding to the `age of infection', i.e.\ the time on a clock that starts when an individual becomes infected. As a key model ingredient we introduce
\begin{equation*}
\pi(\tau,\xi)=\text{the probability that transmission to a susceptible partner happens before $\tau$, given $\xi$}.
\end{equation*}
In the SIR example we have
\begin{equation*}
\pi(\tau,\xi)=1-e^{-\beta\min(\tau,\xi)}.
\end{equation*}
It is important to note a certain asymmetry. On the one hand, there is dependence in the risk of infection of partners of an infectious individual $u$. Their risk of getting infected by $u$ depends on the length of the infectious period of $u$ (and, possibly, other aspects of infectiousness encoded in $\xi$). On the other hand, if $u$ is susceptible, the risk that $u$ itself becomes infected depends on the length of the infectious periods of its various infectious partners. But these partners are independent of one another when it comes to the length of their infectious period (see also~\citet[Section 2.3 `The pitfall of overlooking dependence']{Diekmann2013}). In particular, the probability that an individual escapes infection from its partner, up to at least $\tau$ units of time after the partner became infected, equals
\begin{equation}\label{ref}
\mathcal F(\tau)=1-\int_\Omega \pi(\tau,\xi)m(d\xi).
\end{equation}
For the SIR example~\eqref{ref} boils down to 
\begin{equation}\label{def:F_SIR}
\mathcal F(\tau)=\frac{\gamma}{\beta+\gamma}+\frac{\beta}{\beta+\gamma}e^{-(\beta+\gamma)\tau},
\end{equation}
a formula that can also be understood in terms of two competing events (transmission versus ending of the infectious period) that occur at respective rates $\beta$ and $\gamma$.

As in~\citep{Diekmann1998} and earlier subsections, we consider a static configuration network with uniform degree distribution: every individual is connected to exactly $n$ other individuals. At the end of this section we shall formulate the renewal equation for arbitrary degree distribution. In~\citep{Diekmann1998} an expression for $R_0$ and equations for both final size and the probability of a minor outbreak were derived. In addition, it was sketched how to formulate a nonlinear renewal equation for a scalar quantity, but the procedure is actually that complicated that the resulting equation was \emph{not} written down. 

\begin{figure}[H]
\centering
\includegraphics[scale=0.25]{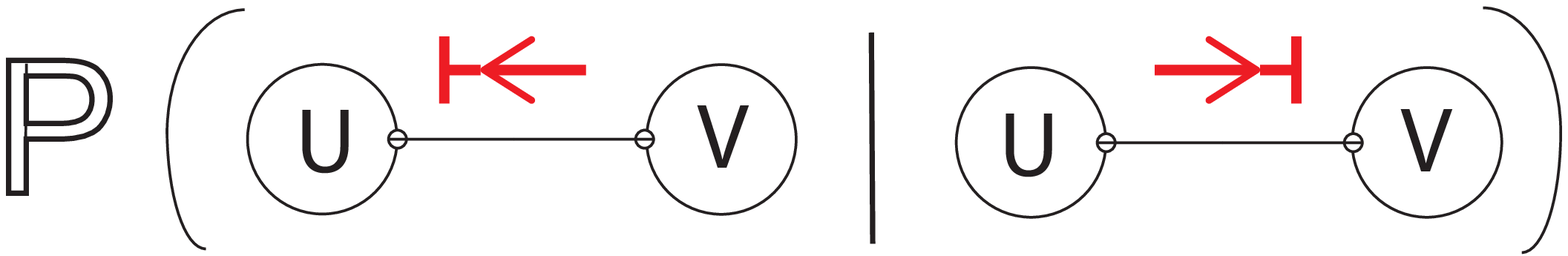}
\caption{\citeauthor{Volz2008} focused on the variable $\theta(t)$ corresponding to the probability that along a randomly chosen partnership between individuals $u$ and $v$ \emph{no} transmission occurred from $v$ to $u$ before time $t$, given that no transmission occurred from $u$ to $v$.}
\label{fig:theta}
\end{figure}

The brilliant idea of~\citet{Volz2008} is to focus on the variable $\theta(t)$ corresponding to the probability that along a randomly chosen partnership between individuals $u$ and $v$ \emph{no} transmission occurred from $v$ to $u$ before time $t$, given that no transmission occurred from $u$ to $v$ (see also Fig.~\ref{fig:theta} for a schematic representation). Here one should think of `probability of transmission' as being defined by $\pi$ (and hence $\mathcal F$) and not require that the individual at the receiving end of the link is indeed susceptible (though, if it actually is, or has been, infectious, the condition of no transmission in the opposite direction is indeed a nontrivial condition).

\begin{figure}[H]
\centering
\includegraphics[scale=0.35]{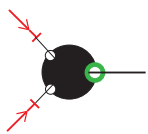}
\caption{Schematic representation of $\bar x$. In this figure, the binding site under consideration is indicated in green. Its owner has three binding sites in total. It is given that no transmission occurs through its other two binding sites.}
\label{fig:xbar}
\end{figure}

The variable $\theta$ corresponds to $\bar x$ introduced in Section~\ref{sec:modelI} and therefore we use that symbol also in this section. We reformulate~\eqref{def:xbarI} as 
\begin{align}
\bar x(t)=\text{prob} &\{\text{a binding site is susceptible at time } t\mid \text{its owner does not become }\nonumber\\
&\qquad\qquad \text{infected through one of its other binding sites before time $t$}\}\label{def:xbar_app}
\end{align}
(see also Fig.~\ref{fig:xbar}). There is an underlying stochastic process in the definition for $\bar x$ that we have not carefully defined here. Yet we shall use the words from the definition to derive a consistency relation that takes the form of a nonlinear renewal equation for $\bar x(t)$. The renewal equation describes the stochastic process starting `far back' in time when all individuals were still susceptible. A precise mathematical definition and an in-depth analysis of the stochastic process can be found in~\citep{Barbour2013}. See~\citep[Sec.~V]{Karrer2010} for a different way of specifying initial conditions. 

To derive the consistency relation for $\bar x(t)$ we shift our focus to the partner that occupies the binding site under consideration. For convenience we call the owner of the binding site under consideration $u$ and the partner that occupies this binding site $v$. Then, given that $u$ does not become infected through one of its $n-1$ other binding sites, $u$ is susceptible at time $t$ if (1) $v$ is susceptible at time $t$ or (2) $v$ is not susceptible at time $t$ but has not transmitted infection to $u$ up to time $t$.  

We begin by determining (1). Given its susceptible partner $u$, individual $v$ is susceptible if its $n-1$ other binding sites are susceptible. Conditioning on its $n-1$ other binding sites not transmitting to $v$, a binding site of $v$ is susceptible at time $t$ with probability $\bar x(t)$. Therefore, given susceptibility of partner $u$, $v$ is susceptible at time $t$ with probability 
\begin{equation}\label{eq:x1consistency}
\bar x(t)^{n-1}.
\end{equation}
This just repeats the consistency relation~\eqref{eq:x1barx} $x_1=\bar x^{n-1}$ stating that the probability $x_1$ that a susceptible binding site is occupied by a susceptible partner is equal to the probability $\bar x^{n-1}$ that a partner of a susceptible individual is susceptible.

Next, suppose that $v$ gets infected at some time $\eta<t$, then $u$ is not infected by $v$ before time $t$ if no transmission occurs in the time interval of length $t-\eta$. The expression~\eqref{eq:x1consistency} has as a corollary that the probability per unit of time that $v$ becomes infected at time $\eta$ equals 
\begin{equation*}
-\frac{d}{d\eta}\big(\bar x(\eta)\big)^{n-1}.
\end{equation*}
Noting that the probability of no transmission to $u$ in the time interval $(\eta, t)$ is $\mathcal F(t-\eta)$ we conclude that necessarily,
\begin{equation}\label{REI}
\bar x(t)=\bar x(t)^{n-1}-\int_{-\infty}^t\left(\frac{d}{d\eta}\big(\bar x(\eta)\big)^{n-1}\right)\mathcal F(t-\eta)d\eta.
\end{equation}
Finally, by integration by parts, we obtain the renewal equation
\begin{equation}\label{RE_special}
\bar x(t)=\mathcal F(\infty)-\int_{-\infty}^t\bar x(\eta)^{n-1}\mathcal F'(t-\eta)d\eta.
\end{equation}
For a configuration network with general degree distribution $(p_n)$ for the number of binding sites $n$ of an individual, exactly the same arguments hold. But now there is randomness of $n$. This leads to the renewal equation (compare with~\eqref{RE_special})
\begin{equation}\label{RE}
\bar x(t)=\mathcal F(\infty)-\int_{-\infty}^t g(\bar x(\eta))\mathcal F'(t-\eta)d\eta,
\end{equation}
with 
\begin{equation*}
g(x)\coloneqq\frac{\sum_{n=1}^\infty n p_n x^{n-1}}{\sum_{m=1}^\infty m p_m}.
\end{equation*}
The solution $\bar x(t)=1$, $-\infty<t<\infty$, of~\eqref{RE} corresponds to the disease free situation. If we put $\bar x(t)=1-h(t)$ and assume $h$ is small, we easily deduce that the linearized equation is given by
\begin{equation}\label{linearRE}
h(t)=-g'(1)\int_{-\infty}^th(\eta)\mathcal F'(t-\eta)d\eta.
\end{equation}
The corresponding Euler-Lotka characteristic equation reads
\begin{equation}\label{charRE}
1=-g'(1)\int_0^\infty e^{-\lambda \tau}\mathcal F'(\tau)d\tau.
\end{equation}
If we evaluate the right hand side of~\eqref{charRE} at $\lambda=0$, we obtain
\begin{equation*}
R_0=g'(1)\big(1-\mathcal F'(\infty )\big)
\end{equation*}
cf.~\citet[eq.\ (12.32), p.\ 294]{Diekmann2013}. In short, the relevant characteristics of the initial phase of an epidemic outbreak are easily obtained from the linearized RE~\eqref{linearRE} (see~\citet{Pellis2015} for a study of the Malthusian parameter, i.e.\ the real root of~\eqref{linearRE}). 

To derive an equation for the final size is even simpler, one takes the limit $t\to\infty$ in~\eqref{RE} to deduce 
\begin{equation}\label{finalRE}
\bar x(\infty)=\mathcal F(\infty)+(1-\mathcal F(\infty)) g(\bar x(\infty)),
\end{equation}
and next observes that the escape probability $s(\infty)$ is given by 
\begin{equation*}
s(\infty)=\sum_{n=1}^\infty p_n\big(\bar x(\infty)\big)^n
\end{equation*}
(to compare to~\citet[eqs.\ (12.36)-(12.38), p.\ 295]{Diekmann2013} identify $\bar q=1-\mathcal F(\infty)$, $\pi=g(\bar x(\infty))$, and rewrite~\eqref{finalRE} as $\pi=g(1-\bar q+\bar q \pi)$).

In the case that $\mathcal F$ is given by~\eqref{def:F_SIR}, the RE
\begin{equation*}
\bar x(t)=\frac{\gamma}{\beta+\gamma}+\beta\int_{-\infty}^t g(\bar x(\eta))e^{-(\beta+\gamma)(t-\eta)}d\eta
\end{equation*}
can be transformed into an ODE for $\bar x$ by differentiation:
\begin{equation*}
\bar x'=\beta g(\bar x)-(\beta+\gamma)\bar x+\gamma.
\end{equation*}
In the special case of Sections~\ref{sec:modelI}-\ref{sec:infI}, we have $p_n=1$ and $p_k=0$ for all $k\neq n$ so $g(x)=x^{n-1}$ and we recover~\eqref{eq:barx}. 

As explained in (O.\ Diekmann, M.\ Gyllenberg, J.A.J.\ Metz. Finite dimensional state representation of linear and nonlinear delay systems. \emph{In preparation}), the natural generalization of~\eqref{def:F_SIR} assumes that $\mathcal F$ is of the form
\begin{equation}\label{generalizationRE}
\mathcal F(\tau)=1-\int_0^\tau \B{\beta}\cdot e^{\eta(\Sigma-\text{diag }\B{\beta})}Vd\eta,
\end{equation}
where, for some $m\in\N$, $\B{\beta}$ and $V$ are non-negative vectors in $\R^m$ while $\Sigma$ is a Positive-Off-Diagonal (POD) $m\times m$ matrix. If $\mathcal F$ is given by~\eqref{generalizationRE}, the variable 
\begin{equation*}
Q(t)\coloneqq\int_{-\infty}^tg(\bar x(\eta))e^{(t-\eta)(\Sigma-\text{diag }\B{\beta})}Vd\eta
\end{equation*}
satisfies the ODE
\begin{equation}\label{eqQ}
\frac{dQ}{dt}=(\Sigma-\text{diag }\B{\beta})Q+g(\bar x)V
\end{equation}
and, since~\eqref{RE} can be rewritten as 
\begin{equation}\label{eqx}
\bar x= \mathcal F(\infty)-\B\beta \cdot Q,
\end{equation}
the equation~\eqref{eqQ} is a closed system once we replace $\bar x$ at the right hand side of~\eqref{eqQ} by the right hand side of~\eqref{eqx}

So one can solve/analyse~\eqref{eqQ} and next use the identity~\eqref{eqx} to draw conclusions about $\bar x$. We conclude that various ODE systems as derived in~\citet{Miller2011} are subsumed in~\eqref{RE} and can be deduced from~\eqref{RE} by a special choice of $\mathcal F$ and differentiation.

\section{Part II: dynamic network without demographic turnover}\label{sec:intermediate}

In Section~\ref{sec:static}, only one environmental variable $\Lambda_-$ is involved in the specification of the dynamics of the susceptible binding sites. In dynamic networks, additional environmental variables play a role. Notably, we have to specify the (probability distribution of the) disease status of a new partner. Before formulating the model for susceptible binding sites, we first consider the network itself in Section~\ref{sec:initialII}. This is needed in order to determine the appropriate `far past' conditions of the susceptible binding site system. 

In Section~\ref{sec:modelII}, the model formulation is divided into three subsections. First, we formulate the model in terms of susceptible binding site probabilities $x$ by following the scheme of five steps presented in Section~\ref{sec:steps}. This allows us to express in terms of $x$ those environmental variables that are defined in terms of susceptible p-level fractions $P_{(-,\B k)}$. We then consider infectious and recovered binding site systems and these allow us to express the other environmental variables in terms of (the history of) $x$ as well.  

\subsection{Network dynamics}\label{sec:initialII}
Binding sites are either free or occupied. We denote the fraction of free binding sites in the population by $F$. We assume that a binding site that is free becomes occupied at rate $\rho F$, while an occupied binding site becomes free at rate $\sigma$. Similar to~\citep{Leung2012} (set $\mu=0$), we find that $F$ satisfies the ODE
\begin{equation*}
\frac{dF}{dt}=-\rho F^2+\sigma(1-F).
\end{equation*}
So we find that $F$ converges to a constant for $t\to\infty$. Therefore, we assume that the fraction of free binding sites is constant, and this constant is again denoted by the symbol $F$. Then $F$ satisfies
\begin{equation}\label{eq:Fb}
F=\frac{\sigma}{\rho F+\sigma}.
\end{equation}
Although we could give an explicit expression in terms of $\sigma$ and $\rho$ for $F$, we prefer to state the more useful identity~\eqref{eq:Fb} that, viewed as an equation, has $F$ as its unique positive root. The network structure, although dynamic, is stable. A randomly chosen binding site (in the pool of all binding sites) is free with probability $F$ and occupied by a partner with probability $1-F$. Later on we shall use that, given that a binding site is free with probability $F$ at time $\tau$, the probability that a binding site is free at time $\xi+\tau$ is $F$ (and the probability that it is occupied at time $\xi+\tau$ is $1-F$).

Finally, later on in Section~\ref{sec:infII}, we also need the probability $\varphi_1(\xi)$ that a binding site is free at time $\xi+\tau$ if it is occupied at time $\tau$. Note that, by the Markov property, this probability only depends on the length $\xi$ of the time interval. Since $\varphi_1(\xi)$ is the unique solution of the initial value problem:
\begin{align*}
\varphi_1'&=-\rho F\varphi_1+\sigma(1-\varphi_1),\\
\varphi_1(0)&=0,
\end{align*}
we have
\begin{equation}\label{varphiII}
\varphi_1(\xi)=\frac{\sigma}{\rho F+\sigma}\left(1-e^{-(\rho F+\sigma)\xi}\right)=F\left(1-e^{-(\rho F+\sigma)\xi}\right)
\end{equation}
where we used~\eqref{eq:Fb} in the second equality.

\subsection{Model formulation}\label{sec:modelII}
\subsubsection{Susceptibles}\label{sec:suscII}
We describe the dynamics of susceptible binding sites in terms of $x$-probabilities. Consider a susceptible binding site and suppose its owner does not become infected through one of its other $n-1$ binding sites for the period under consideration. An occupied binding site (in states 1, 2, or 3) becomes free if it loses its partner (corresponding to a transition to state 0). This occurs at rate $\sigma$. A binding site that is free, i.e.\ a binding site in state 0, can acquire a partner. The rate at which this occurs is $\rho F$ where $F$ is the fraction of free binding sites defined by~\eqref{eq:Fb}. Free binding sites either have a susceptible, infectious, or recovered owner. So there are three additional environmental variables, viz.\ the fraction of binding sites that are free and have disease status $d$ (i.e.\ having owners with disease status $d$), we denote these by $F_d$, $d\in\{-,+,\ast\}$. Then $F=F_-+F_++F_\ast$. Finally, there are infection and recovery events that can cause state transitions (as in the case of a static network in Section~\ref{sec:static}).

Long ago in time, by assumption, all individuals (and therefore binding sites) are susceptible. In accordance with Section~\ref{sec:initialII} the fraction of free and susceptible binding sites is equal to $F$ and the fraction of susceptible binding sites occupied by susceptible partners is equal to $1-F$, i.e.\ we have `far past' conditions
\begin{align}\label{eq:II_initial}
x_0(-\infty)&=F,\quad x_1(-\infty)=1-F,\quad x_2(-\infty)=0=x_3(-\infty).
\end{align}

Let $\B F=(F_-,F_+,F_\ast)$. The environmental variables $\B F$ and $\Lambda_-$ are p-level quantities that we have yet to specify. Putting together the various assumptions described above, the dynamics of $x$ is governed by the system:
\begin{align}\label{eq:ODExII}
\frac{dx(t)}{dt}&=M\big(\B{F}(t),\Lambda_-(t)\big)x(t),
\end{align}
with `far past' conditions~\eqref{eq:II_initial}, and
\begin{align}\label{eq:MII}
M(\B{F},\Lambda_-)&=\begin{pmatrix}-\rho F &\sigma &\sigma &\sigma\\\rho F_-& -(\beta\Lambda_-+\sigma) &0&0\\\rho F_+&\beta\Lambda_-&-(\beta+\sigma+\gamma)&0\\\rho F_\ast&0&\gamma&-\sigma\end{pmatrix}.
\end{align}

Next, in \emph{step 2}, we define the environmental variables in terms of p-level fractions. The definition~\eqref{def:Lambda-} of $\Lambda_-$ in terms of p-level fractions carries over. We define the fractions of free binding sites in terms of p-level fractions as follows:
\begin{equation}\label{def:F}
F_d(t)=\frac1n\sum_{\B{k}}(n-k_1-k_2-k_3)P_{(d,\B{k})}(t),
\end{equation}
where the sum is over all possible configurations of $\B k$ with $0\leq k_1+k_2+k_3\leq n$. 

In \emph{step 3} we define the i-level probabilities $p_{(-,\B k)}(t)$ in terms of the $x$ probabilities by using the conditional independence of binding sites:
\begin{equation}\label{def:x2pII}
p_{(-,\B k)}(t)=\frac{n!}{k_0!\, k_1!\, k_2!\,k_3!}\left(x_0^{k_0}x_1^{k_1}x_2^{k_2}x_3^{k_3}\right)(t).
\end{equation}

As in the static network case I, the i-level probabilities coincide with the p-level fractions, i.e.~\eqref{eq:i2p} holds. This is \emph{step 4} in our model formulation.

Then, in \emph{step 5}, we can express the environmental variables $\Lambda_-$ and $F_-$ in terms of $x$-probabilities. By combining~\eqref{def:x2pII} with~\eqref{eq:i2p} and~\eqref{def:Lambda-}, we again find~\eqref{eq:Lambda-I} to hold (only now the $x$ are defined by the system of ODE~\eqref{eq:ODExII}). By combining~\eqref{def:x2pII} with~\eqref{eq:i2p} and~\eqref{def:F} we find that
\begin{equation}\label{eq:F0_intermediate}
F_-(t)=x_0(t)\bar x(t)^{n-1},
\end{equation}
exactly as the interpretations of $F_-$ and $x$ would suggest.

Before we can specify $F_+$ and $F_\ast$ in terms of (the history of) $x$ we need to define p-level fractions $P_{(+, \B k)}(t)$ and $P_{(\ast, \B k)}(t)$. We do so in the next section where we turn to infectious and recovered binding site systems.

\subsubsection{After suscetibility is lost}\label{sec:infII}
If an individual becomes infected at time $t_+$, the binding site through which infection is transmitted is from that point on `exceptional'. Then, given that its owner became infected at time $t_+$ and that it does not recover for the time under consideration, we consider an infectious binding site. Let $y_i^\text{e}(t\mid t_+)$ denote the probability that the exceptional binding site is in state $i$ at time $t$ and $y_i(t\mid t_+)$ this same probability for a non-exceptional binding site. 

As in Section~\ref{sec:infI}, the exceptionalness plays a role only in the states at epidemiological birth, i.e.\ at time $t_+$. The exceptional binding site is with probability one in state 2 at time $t_+$. The states of all other binding sites are distributed according to $x(t_+)/\bar x(t_+)$. Therefore, we put boundary conditions
\begin{align*}
y_0^\text{e}(t_+\mid t_+)=0,\qquad &y_0(t_+\mid t_+)=x_0(t_+)/\bar x(t_+),\\
y_1^\text{e}(t_+\mid t_+)=0,\qquad &y_1(t_+\mid t_+)=x_1(t_+)/\bar x(t_+),\\
y^\text{e}_2(t_+\mid t_+)=1,\qquad &y_2(t_+\mid t_+)=x_2(t_+)/\bar x(t_+),\\
y^\text{e}_3(t_+\mid t_+)=0,\qquad &y_3(t_+\mid t_+)=x_3(t_+)/\bar x(t_+). 
\end{align*}
Since the individual does not recover in the period under consideration, the infectious binding sites behave independently of one another. 

The dynamics of $y^\text{e}$ and $y$ are both governed by the system
\begin{equation}\label{ODE:yII}
\begin{aligned}
%\frac{d y^\text{e}(t\mid t_+)}{d t}&=M_+\big(\B{F}(t),\Lambda_+(t)\big)y^\text{e}(t\mid t_+),\\
\frac{d y(t\mid t_+)}{d t}&=M_+\big(\B{F}(t),\Lambda_+(t)\big)y(t\mid t_+),
\end{aligned}
\end{equation}
with \begin{align*}
M_+(\B{F},\Lambda_+)&=\begin{pmatrix}-\rho F &\sigma &\sigma &\sigma\\\rho F_-& -(\beta\Lambda_++\sigma) &0&0\\\rho F_+&\beta\Lambda_+&-(\sigma+\gamma)&0\\\rho F_\ast&0&\gamma&-\sigma\end{pmatrix},
\end{align*}
Note that there is no rate $\gamma$ of leaving the infectious state as we condition on the owner remaining infectious in the period under consideration. Furthermore, note that, contrary to the network case I of Section~\ref{sec:infI}, the exceptional binding site can lose its epidemiological parent by separation. Therefore $y_1^\text{e}(t\mid t_+)>0$ for $t> t_+$. 

Next, similarly to Section~\ref{sec:infI}, by combinatorics (but now probabilities $y_0^\text{e}$ and $y_1^\text{e}$ are not equal to zero for $t\geq t_+$), we find that the probability $\phi_{(+,\B k)}(t\mid t_+)$ that an individual, infected at time $t_+$, is in state $(+,\B k)$ at time $t\geq t_+$ is given by
\begin{align}
\phi_{(+,\B k)}(t\mid t_+)=&\frac{n!}{k_0!\, k_1!\,k_2!\,k_3!}\left(\frac{k_0}n\,y_0^\text{e}\ y_0^{k_0-1}y_1^{k_1}y_2^{k_2}y_3^{k_3}+\frac{k_1}n\,y_1^\text{e}\ y_0^{k_0}y_1^{k_1-1}y_2^{k_2}y_3^{k_3}\right.+\nonumber\\
&\quad \left.\frac{k_2}n\,y_2^\text{e}\ y_0^{k_0}y_1^{k_1}y_2^{k_2-1}y_3^{k_3}+\frac{k_3}n\, y_3^\text{e}\ y_0^{k_0}y_1^{k_1}y_2^{k_2}y_3^{k_3-1}\right)(t\mid t_+).\label{def:inf_pb}
\end{align}

The probability $P_{(+,\B k)}(t)$ that a randomly chosen individual is in state $(+, \B k)$ at time $t$ is obtained by taking into account the time of infection $t_+$ and the probability~\eqref{eq:no_rec} that an individual has not recovered time $t-t_+$ after infection. The definition~\eqref{eq:incidence} for the incidence carries over (but now with the $x$ defined by the ODE system~\eqref{eq:ODExII}). So
\begin{equation}\label{def:inf_Ppb}
P_{(+,\B k)}(t)=p_{(+,\B k)}(t)=\int_{-\infty}^te^{-\gamma(t-t_+)}\beta nx_2\bar x^{n-1}(t_+)\phi_{(+,\B k)}(t\mid t_+)dt_+.
\end{equation}
By combining~\eqref{eq:i2p} and the expression~\eqref{def:inf_Ppb} for $P_{(+\B k)}$ in terms of $y$ and $y^\text{e}$ we can redefine $F_+$ in terms of the history of $x$ as we will show now. First of all, combining~\eqref{def:inf_pb},~\eqref{def:inf_Ppb} and~\eqref{def:F} we express $F_+$ in terms of $y$ and $y^\text{e}$:
\begin{equation}\label{eq:F+yb}
F_+(t)=\frac1n\int_{-\infty}^t e^{-\gamma(t-t_+)}\beta nx_2\bar x^{n-1}(t_+)\left(y_0^\text{e}\, \bar y^{n-1}+(n-1)\bar y^\text{e}\, y_0\, \bar y^{n-2}\right)(t\mid t_+)dt_+.
\end{equation}
where $\bar y=y_0+y_1+y_2+y_3$ and $\bar y^\text{e}=y_0^\text{e}+y_1^\text{e}+y_2^\text{e}+y_3^\text{e}$. Since $y^\text{e}$ and $y$ are probability vectors, 
\begin{equation*}
\bar y^\text{e}(t\mid t_+)=1=\bar y(t\mid t_+).
\end{equation*} 
Next, we consider the probabilities $y_0, y_0^\text{e}$. Note that 
\begin{equation*}
y_0^\text{e}(t\mid t_+)=\varphi_1(t-t_+),\label{eq:y0_exc}\\
\end{equation*}
with $\varphi_1$ given by~\eqref{varphiII}. The dynamics of $y_0$ are described in terms of $y_0$ and the history of $x_0/\bar x$ (by means of the boundary condition). We can solve for $y_0$. This yields
\begin{equation*}
y_0(t\mid t_+)=\varphi_1(t-t_+)+\frac{x_0}{\bar x}(t_+)e^{-(\rho F+\sigma)(t-t_+)}
\end{equation*}
(note that time of infection $t_+$ matters in this probability and not only the length $t-t_+$ of the time interval). We can further simplify~\eqref{eq:F+yb} to obtain 
\begin{align}
F_+(t)%&=\frac1n\int_{-\infty}^te^{-\gamma(t-t_+)}\beta n x_2\bar x^{n-1}(t_+)\left(\varphi_1(t-t_+)+(n-1)y_0(t\mid t_+)\right)dt_+\nonumber\\
&=\frac1n\int_{-\infty}^t e^{-\gamma(t-t_+)}\beta n x_2\bar x^{n-1}(t_+)\nonumber\\
&\qquad\quad\left\{\varphi_1(t-t_+)+(n-1)\left(\varphi_1(t-t_+)+\frac{x_0}{\bar x}(t_+)e^{-(\rho F+\sigma)(t-t_+)}\right)\right\}dt_+\label{def:F+b_bs},
\end{align}
which only depends on the model parameters and past probabilities $x_i$ for susceptible binding sites.

We can use the consistency condition for the total fraction of free binding sites:
\begin{equation}\label{eq:F3II}
F_\ast(t)=F-F_-(t)-F_+(t).
\end{equation}
to express $F_\ast$ in terms of the history of $x$ (use~\eqref{eq:F0_intermediate} for $F_-$ and~\eqref{def:F+b_bs} for $F_+$). So this specifies all environmental variables for the susceptible binding site system $x$ in terms of (the history of $x$). 

Next, similar to case I of Section~\ref{sec:infI}, we consider recovered individuals and their binding sites. Suppose that the infectious individual, that was infected at time $t_+$, recovers at time $t_\ast$. After recovery, we still distinguish between the exceptional binding site and the $n-1$ other binding sites. We introduce probabilities $z_i^\text{e}(t\mid t_+, t_\ast)$ and $z_i(t\mid t_+, t_\ast)$ for recovered binding sites. The $y$ and $y^\text{e}$ probabilities yield the conditions for $z$ and $z^\text{e}$ at time $t=t_\ast$, i.e.
\begin{alignat*}{3}
z_0^\text{e}(t_\ast\mid t_+, t_\ast)&=y_1^\text{e}(t_\ast\mid t_+),\qquad &z_0(t_\ast\mid t_+, t_\ast)&=y_1(t_\ast\mid t_+),\\
z_1^\text{e}(t_\ast\mid t_+, t_\ast)&=y_1^\text{e}(t_\ast\mid t_+),\qquad &z_1(t_\ast\mid t_+, t_\ast)&=y_1(t_\ast\mid t_+),\\
z^\text{e}_2(t_\ast\mid t_+, t_\ast)&=y_2^\text{e}(t_\ast\mid t_+),\qquad &z_2(t_\ast\mid t_+, t_\ast)&=y_2(t_\ast\mid t_+),\\
z^\text{e}_3(t_\ast\mid t_+, t_\ast)&=y_3^\text{e}(t_\ast\mid t_+),\qquad &z_3(t_\ast\mid t_+, t_\ast)&=y_3(t_\ast\mid t_+).
\end{alignat*}
The dynamics of $z$ and $z^\text{e}$ are described by the system of ODE for $y$ and $y^\text{e}$, with the mean field at distance one quantity $\Lambda_+$ replaced by $\Lambda_\ast$ where $\Lambda_\ast$ is defined in terms of p-level fractions by~\eqref{def:Lambda*} and hence is given by~\eqref{def:Lambda*x} in terms of $x$-probabilities.

Let $\psi_{(\ast, \B k)}(t\mid t_+, t_\ast)$ denote the probability that a recovered individual is in state $(\ast,\B k)$ given that it was infected at time $t_+$ and recovered at time $t_\ast$. Then $\psi_{(\ast,\B k)}(t\mid t_+,t_\ast)$ can be expressed in terms of $z$ and $z^\text{e}$ by replacing $\phi$ in~\eqref{def:inf_pb} by $\psi$, $y_i$ by $z_i$, and $y_i^\text{e}$ by $z_i^\text{e}$.

The unconditional probability $p_{(\ast, \B k)}(t)$ is then obtained by taking into account time of infection $t_+$ and recovery time $t_\ast$:
\begin{equation}\label{def:rec_Ppb}
p_{(\ast,\B k)}(t)=\int_{-\infty}^t\int_{-\infty}^{t_\ast}\gamma e^{-\gamma(t_\ast-t_+)}\beta n x_2\bar x^{n-1}(t_+)\psi_{(\ast,\B k)}(t\mid t_+, t_\ast)dt_+dt_\ast,
\end{equation}
which, by relation~\eqref{eq:i2p}, is equal to the p-level fraction $P_{(\ast,\B k)}(t)$. Note that we can also use this definition of $P_{(\ast, \B k)}(t)$ to define $F_\ast$ in terms of $x$ similar to the way we did for $F_+$ in~\eqref{def:F+b_bs}.

\subsubsection{One renewal equation or a system of six ODE, whatever you like}\label{sec:REII}
We ended the model formulation in Section~\ref{sec:suscII} by defining the environmental variables $\Lambda_-$ and $F_-$ in terms of $x$ (eqs.~\eqref{eq:Lambda-I} and~\eqref{eq:F0_intermediate}). Subsequently, in Section~\ref{sec:infII}, by considering infectious binding site probabilities $y$, and $y^\text{e}$, we also defined $F_+$ and $F_\ast$ in terms of $x$ (eqs.~\eqref{def:F+b_bs} and~\eqref{eq:F3II}). Combining these formulas, we find that the system describing the dynamics of susceptible binding sites is given by:
\begin{equation}\label{eq:dynamic1ODEx}
\begin{aligned}
x_0'&=-\rho F x_0+\sigma(x_1+x_2+x_3)\\
x_1'&=\rho x_0^2\bar x^{n-1}-\left(\sigma+\beta (n-1)\frac{x_2}{\bar x}\right)x_1\\
x_2'&=\rho F_+x_0+\beta (n-1)\frac{x_2}{\bar x}x_1-(\sigma+\beta+\gamma)x_2\\
x_3'&=\rho \big(F-x_0\bar x^{n-1}-F_+\big)x_0+\gamma x_2-\sigma x_3,
\end{aligned}
\end{equation}
with $F_+$ given by~\eqref{def:F+b_bs} and with `far past' condition
\begin{equation}\label{eq:initialODEx}
x_0(-\infty)=F,\quad x_1(-\infty)=1-F,\quad x_2(-\infty)=0=x_3(-\infty).
\end{equation}

The ODE~\eqref{eq:dynamic1ODEx} for $x$ together with the expression~\eqref{def:F+b_bs} for $F_+$ yields a closed system of five equations. By substituting expression~\eqref{def:F+b_bs} in system~\eqref{eq:dynamic1ODEx}, one can view~\eqref{eq:dynamic1ODEx} as a system of four delay differential equations. The dynamics of the $1/6(n+1)(n+2)(n+3)$ i-level probabilities (hence p-level fractions) $p_{(-,\B k)}$ for susceptible individuals are fully determined by this set of four delay differential equations (regardless of $n$). 

Alternatively, we can view the solution $x(t)$ of~\eqref{eq:dynamic1ODEx}-\eqref{eq:initialODEx} as fully determined by $F_+|_{(-\infty,t]}$. Interpreting $x_2$, $\bar x$, and $x_0$ at the right hand side of~\eqref{def:F+b_bs} in this manner, we arrive at the conclusion that the dynamics are fully determined by a single renewal equation for $F_+$. 

One may prefer a system consisting only of ODE rather than a delay system. We can in fact reason directly in terms of the interpretation to derive an ODE for $F_+$. In order to do so, we first consider the fraction $I(t)=\sum_{\B k} P_{(+,\B k)}$ of infecteds in the population. This fraction decreases when infecteds recover. Infecteds recover at a constant rate $\gamma$. The fraction $I$ increases when a susceptible individual becomes infected so there is the positive term~\eqref{eq:incidence} in the ODE for $I$ (combine~\eqref{def:x2pII} with~\eqref{eq:i2p} and~\eqref{eq:incidence}, the $x$ are defined by the ODE system~\eqref{eq:ODExII})). We find that the dynamics of $I$ are described by the following ODE:
\begin{equation}\label{eq:ODE_inf_II}
\frac{dI}{dt}=\beta n x_2\bar x^{n-1}-\gamma I,
\end{equation}
with `far past' condition $I(-\infty)=0$. Next, we consider $F_+$. Any infectious owner recovers at constant rate $\gamma$. In addition, partnership formation and separation affect the fraction of free infectious binding sites. There is a rate $\rho F$ at which free binding sites become occupied. The fraction of infectious binding sites that are occupied is given by $I-F_+$ and the rate at which these binding sites become free is $\sigma$. Then, finally, a susceptible individual with $k_2$ infectious partners becomes infected at rate $\beta k_2$, taking into account all $0\leq k_2\leq n$ we find probability per unit of time $\beta n {x_2}\bar x^{n-1}$ at which a susceptible individual becomes infected. The probability that a non-exceptional binding site is free and susceptible upon infection is $x_0/\bar x$, so the expected fraction of free binding sites created upon infection of a susceptible individual is $\frac1n(n-1)x_0/\bar x$. Hence there is a flow $\beta(n-1) x_0x_2\bar x^{n-2}$ into $F_+$. We have the following ODE for $F_+$:
\begin{equation}\label{eq:ODE_F+_II}
\frac{dF_+}{dt}=\beta(n-1)x_0x_2\bar x^{n-2}-(\rho F+\gamma)F_++\sigma(I-F_+),
\end{equation}
with `far past' condition $F_+(-\infty)=0$. 

Alternatively, we can derive the ODE~\eqref{eq:ODE_F+_II} for $F_+$ by differentiating~\eqref{def:F+b_bs} with respect to $t$. Note that we can express $I$ in terms of $x$ by first expressing it in terms of $y$ and $y^\text{e}$ (similar to $F_+$ in Section~\ref{sec:infII}). This yields $I(t)=\int_{-\infty}^te^{-\gamma(t-t_+)}\beta n x_2\bar x^{n-1}(t_+)dt_+$. 

The combination of~\eqref{eq:dynamic1ODEx} with~\eqref{eq:ODE_inf_II} and~\eqref{eq:ODE_F+_II} yields a six-dimensional closed system of ODE. (Compare with the slightly different but related network model called the `dormant contacts' model of~\citep{Miller2011}. Presumably~\eqref{eq:dynamic1ODEx},~\eqref{eq:ODE_inf_II},~\eqref{eq:ODE_F+_II} is a transformed but equivalent version of their system (3.11)-(3.16).)

Both~\eqref{def:F+b_bs} and~\eqref{eq:dynamic1ODEx}-\eqref{eq:ODE_F+_II} can be used to represent the system. In terms of the number of equations, it does not matter too much which system one considers. In the first case, one renewal equation is needed compared to six ODE in the second case. In both formulations one can determine $r$ and $R_0$ with not too much effort. In Section~\ref{sec:R0II} below, we will use a pragmatic mixture. This gives us a way of determining $r$ and $R_0$ that prepares for the characterization of $r$ and $R_0$ in case III in Section~\ref{sec:R0III} (where a model formulation in terms of only ODE becomes troublesome).

\subsection{The beginning and end of an epidemic: $R_0$, $r$ and final size}\label{sec:R0II}
First, just as in case I, the final size is given by
\begin{equation*}
1-\bar x(\infty)^n.
\end{equation*}
But while in case I we derived a simple scalar equation for $\bar x(\infty)$ (\eqref{eq:app_finalsize} or~\eqref{finalRE}), depending explicitly on the parameters, we did not, despite fanatical efforts, manage to derive such an equation from the implicit characterization by~\eqref{eq:dynamic1ODEx},~\eqref{eq:ODE_inf_II},~\eqref{eq:ODE_F+_II}; see also Appendix~\ref{initial}.

Next, in the rest of this section, we use the binding site level system~\eqref{eq:dynamic1ODEx}-\eqref{eq:ODE_F+_II} to consider the beginning of an epidemic and determine $R_0$ and $r$. The point here is not only to use~\eqref{eq:dynamic1ODEx}-\eqref{eq:ODE_F+_II} to find threshold parameters but to find threshold parameters with their usual interpretation of $R_0$ and $r$. %\red{The second part of this section is devoted to the (rather unsuccesful attempt at a) characterization of the final size. }

Using the same arguments as in network case I of Section~\ref{sec:static}, we find that a threshold parameter for the disease free steady state of system~\eqref{eq:dynamic1ODEx}-\eqref{eq:ODE_F+_II} on the binding site level is also a threshold parameter for the disease free steady state of the p-level system. 

The disease free steady state of~\eqref{eq:dynamic1ODEx} is given by $\tilde x_0=F$, $\tilde x_1=1-F$, $\tilde x_2=0=\tilde x_3$. Linearization in this state yields a decoupled system of equations for the linearized $x_2$ and $F_+$ equations. We let $\hat x_2$ and $\hat F_+$ denote the variables in the linearization in the disease free steady state. Note that, in the disease free steady state $\tilde y_0(t\mid t_+)=\varphi_1(t-t_+)+Fe^{-(\rho F+\sigma)(t-t_+)}=F$, i.e.\ in the disease free steady state, the probability that an infectious binding site is free at time $t$ given that it is free at time $t_+$ is equal to the probability $F$ that a randomly chosen binding site is free. Then
\begin{align}
\hat x_2'&=\rho F \hat F_++\beta(n-1)(1-F)\hat x_2-(\sigma+\beta+\gamma)\hat x_2\nonumber\\
\hat F_+(t)&=\int_0^\infty e^{-\gamma\xi}\beta \hat x_2(t-\xi)\big(\varphi_1(\xi)+(n-1)F\big)d\xi\label{eq:Flinear}
\end{align}
which can be viewed as a linear delay differential equation for $\hat x_2$. In order to obtain an informative version of the corresponding characteristic equation, we rewrite it as a renewal equation for $\hat x_2$.

Variation of constants for the ODE for $\hat x_2$ yields:
\begin{align*}
\hat x_2(t)&=\int_{-\infty}^te^{-(\sigma+\beta+\gamma)(t-\xi)}\left(\rho F \hat F_+(\xi)+\beta(n-1)(1-F)\hat x_2(\xi)\right)d\xi\\
&=\int_0^\infty e^{-(\sigma+\beta+\gamma)\xi}\left(\rho F \hat F_+(t-\xi)+\beta(n-1)(1-F)\hat x_2(t-\xi)\right)d\xi.
\end{align*}
Substituting~\eqref{eq:Flinear} into this expression yields the renewal equation for $\hat x_2$:
\begin{align*}
\hat x_2(t)&=\int_0^\infty e^{-(\sigma+\beta+\gamma)\xi}\bigg\{\rho F\int_0^\infty e^{-\gamma\eta}\beta \hat x_2(t-\xi-\eta)\\
&\phantom{=\ \qquad}\big(\varphi_1(\eta)+(n-1)F\big)d\eta+\beta(n-1)(1-F)\hat x_2(t-\xi)\bigg\}d\xi\\
&=\int_0^\infty \hat x_2(t-\xi)k(\xi)d\xi,
\end{align*} 
with
\begin{align*}
k(\xi)&=\beta e^{-(\sigma+\beta+\gamma)\xi}(n-1)(1-F)\\
&\phantom{=\ }+\int_{0}^\xi \beta e^{-(\sigma+\beta+\gamma)\eta}e^{-\gamma(\xi-\eta)}\rho F \big(\varphi_1(\xi-\eta)+(n-1)F\big)d\eta
\end{align*}
(where the rearrangement of the terms in the integrals is in preparation for the interpretation). Next, we substitute the ansatz $\hat x_2(t)=e^{\lambda t}$, and obtain the characteristic equation
\begin{equation}\label{char_eq}
1=\int_0^\infty e^{-\lambda \xi}k(\xi)d\xi,
\end{equation}
There is a unique real root to~\eqref{char_eq} and this root is by definition the Malthusian parameter $r$. We define $R_0=\int_0^\infty k(\xi)d\xi$. Then sign($r$)=sign($R_0-1$), and we find that $R_0$ is a threshold parameter with threshold value one for the stability of the disease free steady state, with $R_0=\int_0^\infty k(\xi)d\xi$ equal to
\begin{align}
&\int_0^\infty\beta e^{-(\sigma+\beta+\gamma)\xi}(n-1)(1-F)d\xi\nonumber\\
\phantom{=\ }&+\int_0^\infty\int_{0}^{\xi}\beta e^{-(\sigma+\beta+\gamma)\eta}e^{-\gamma(\xi-\eta)}\rho F\big(\varphi_1(\xi-\eta)+(n-1)F\big) d\eta \,d\xi\nonumber\\
=&\int_0^\infty\beta e^{-(\sigma+\beta+\gamma)\xi}d\xi\,\bigg\{(n-1)(1-F)+\int_{0}^{\infty}e^{-\gamma\tau}\rho F\big(\varphi_1(\tau)+(n-1)F\big)d\tau\bigg\}\label{meannumber}
\end{align}
We can evaluate the integrals and find an explicit expression for $R_0$. However, the interpretation is easier in the form it is written now. 

First of all, consider a newly infected individual $u$. Individual $u$ transmits infection to a susceptible partner with probability $\int_0^\infty\beta e^{-(\sigma+\beta+\gamma)\xi}d\xi={\beta}/({\beta+\sigma+\gamma})$. By multiplying this probability with the expected number of susceptible partners $u$ has at epidemiological birth plus the expected number of susceptible partners $u$ acquires during its infectious period after epidemiological birth, we obtain $R_0$. As we will explain now, these are exactly the two terms in $\{\cdots\}$ of~\eqref{meannumber}.

The mean number of susceptible partners of $u$ \emph{at} epidemiological birth is $(n-1)(1-F)$ (note that, in addition to the susceptible partners, $u$ has $(n-1)F$ free and 1 exceptional binding site). This is the first term in $\{\cdots\}$ of~\eqref{meannumber}. We are left with determining the expected number of susceptible partners $u$ acquires \emph{after} epidemiological birth. This goes as follows. At time $\tau$ after $u$ became infected, $u$ has not recovered yet with probability $e^{-\gamma\tau}$. The exceptional binding site of $u$ is free at time $\tau$ with probability $\varphi_1(\tau)$ (see~\eqref{varphiII}). Each of the $n-1$ non-exceptional binding sites of $u$ are free with probability $F$ regardless of whether they were free or occupied at epidemiological birth (recall Section~\ref{sec:initialII}). Note that a free binding site becomes occupied by a susceptible partner at rate $\rho F$ (at the beginning of the epidemic). Integrating over all possible lengths $\tau>0$ of the infectious period, we find that $\int_{0}^{\infty}e^{-\gamma\tau}\rho F\big(\varphi_1(\tau)+(n-1)F\big)d\tau$ is the expected number of additional susceptible partners of $u$ in its infectious period after epidemiological birth. 

Note that we made the distinction of the susceptible partners \emph{at} and \emph{after} epidemiological birth of $u$ but what really matters is the \emph{total} number of susceptible partners in the infectious period of $u$. So really, we did not need to make any distinction between at and after epidemiological birth. But this distinction is essential in Section~\ref{sec:R0III} of case III. The distinction here serves both to illustrate this difference with case III and as a preparation for case III.

Finally, in the same spirit, we would like to mention that rather than taking the perspective of an infectious individual/binding site, we can also take the perspective of a susceptible binding sites `at risk' of infection, i.e.\ susceptible binding sites occupied by infectious partners, and interpret $R_0$ in that way. In the present context this does not change much. Therefore we refrain from elaborating. We leave this for Section~\ref{sec:R0III} of case III where this different perspective leads to a major simplification compared to the `standard' perspective of infectious binding sites that we took here.

\section{Part III: dynamic network with demography}\label{sec:demography}
In this part, the network is not only dynamic due to partnership formation and separation but also due to demographic turnover. We assume that there is a constant per capita death rate $\mu$ and a constant population birth rate so that the population size is in equilibrium and the age of individuals is exponentially distributed with parameter $\mu$. At birth, an individual does not have any partners. Details are presented in~\citep{Leung2012}. 

\subsection{Network dynamics}
In a world with demographic turnover, next to calendar time, also age matters. We keep track of both age $a$ and time of birth $t_b$ of an individual (calendar time is then given by $t=a+t_b$). When we speak about the age and time of birth of a binding site, we mean the age and time of birth of its owner. Often, we assume that the owner of a binding site does not die in the period under consideration. By assumption, at age zero, a binding site is free. A free binding site becomes occupied at rate $\rho F$ where $F$ denotes the total fraction of free binding sites in the population. This $F$ is assumed to be constant (see~\citep{Leung2012} for the justification) and satisfies
\begin{equation}\label{eq:F_III}
F=\frac{\sigma+2\mu}{\rho F+\sigma+2\mu}
\end{equation}
(compare with~\eqref{eq:Fb}). If the binding site is occupied, then it becomes free at rate $\sigma+\mu$ where $\sigma$ and $\mu$ represent separation and death of partner, respectively. 

In this section we will also make use of the following binding site probabilities (where, as usual, we condition on the owner not dying in the period under consideration). We let $\varphi_0(a)$ denote the probability that a binding site is free at age $a+\alpha$, given that it was free at age $\alpha$, and $\varphi_1(a)$ denotes the probability that a binding site is free at age $a+\alpha$, given that it is occupied at age $\alpha$. Note that, by the Markov property, these probabilities only depend on the time interval $a$ (recall that $F$ is constant). The dynamics of $\varphi_i$ as a function of $a$ is described by
\begin{equation*}
\frac{d\varphi_i}{da}=-\rho F \varphi_i+(\sigma+\mu)(1-\varphi_i),
\end{equation*} 
with initial conditions, respectively,
\begin{equation*}
\varphi_0(0)=1,\qquad \varphi_1(0)=0.
\end{equation*}
The explicit expressions for the $\varphi_i$ are given by
\begin{align}
\varphi_0(a)&=\frac{\sigma+\mu}{\rho F+\sigma+\mu}+\frac{\rho F}{\rho F+\sigma+\mu}e^{-(\rho F+\sigma+\mu)a},\label{eq:epsilon0III}\\
\varphi_1(a)&=\frac{\sigma+\mu}{\rho F+\sigma+\mu}\left(1-e^{-(\rho F+\sigma+\mu)a}\right).\label{eq:epsilon1III}
\end{align}
See also~\citep[eq.~(10)]{Leung2012} (where $\epsilon(a)$ can be identified with $1-\varphi_0(a)$) and~\citep[eq.~(67)]{Leung2015a} (where $\epsilon_0(t)$ and $\epsilon_1(t)$ can be identified with $1-\varphi_0(t)$ and $1-\varphi_1(t)$, respectively). 

Furthermore, we have the identity
\begin{equation}\label{eq:Fvarphi_0}
F=\int_0^\infty \mu e^{-\mu a}\varphi_0(a)da
\end{equation}
(use~\eqref{eq:F_III}), expressing that a randomly chosen binding site is free with probability $F$. So, according to Bayes' Theorem, the probability density function of the age of (the owner of) a free binding site is given by
\begin{equation}\label{eq:densityIII}
\pi_0(a)=\frac{\mu e^{-\mu a}\varphi_0(a)}{F}.
\end{equation}
Similarly, the probability density function of the age of (the owner of) a randomly chosen occupied binding site is
\begin{equation}\label{eq:densityIII2}
\pi_1(a)=\frac{\mu e^{-\mu a}\big(1-\varphi_0(a)\big)}{1-F}.
\end{equation}
(in view of the derivation of a formula for $R_0$ in Section~\ref{sec:R0III} below, we remark that $\pi_0$ and $\pi_1$ should be compared to probability distributions $q$ and $Q$, respectively, in~\citep{Leung2015a}; the difference is that $q$ and $Q$ concern the number of partners while $\pi_0$ and $\pi_1$ concern the age; the probability distributions, however, provide the same information).

\subsection{Model formulation}

\subsubsection{Susceptibles}\label{sec:suscIII}
Demography does not give rise to any additional environmental variables, we still deal with the mean field at distance one variable $\Lambda_-$, and the fractions $F_d$ of free binding sites with disease status $d$, $d\in\{-,+,\ast\}$. 

We follow the steps 1-5 of Section~\ref{sec:steps}. In \emph{step 1} we consider $x$-probabilities. Consider a susceptible binding site, born at time $t_b$, and suppose that its owner, for the period under consideration, does not die and does not become infected through one of its other $n-1$ binding sites. The dynamics of $x$ as a function of age are described by the following system of equations:
\begin{align}\label{ODE_III_1}
\frac{d x(a\mid t_b)}{d a}&=M\big(\B{F}(t_b+a),\Lambda_-(t_b+a)\big)x(a\mid t_b),
\end{align}
with
\begin{align}\label{ODE_III_2}
M(\B{F},\Lambda_-)&=\begin{pmatrix}-\rho F &\sigma+\mu &\sigma+\mu &\sigma+\mu\\\rho F_-& -(\beta\Lambda_-+\sigma+\mu) &0&0\\\rho F_+&\beta\Lambda_-&-(\beta+\sigma+\mu+\gamma)&0\\\rho F_\ast&0&\gamma&-(\sigma+\mu)\end{pmatrix}.
\end{align}
An individual is assumed to be susceptible without any partners at birth (and therefore the same applies to all its binding sites). So we have the birth conditions 
\begin{align}\label{ODE_III_3}
x_0(0\mid t_b)&=1,\quad x_1(0\mid t_b)=0=x_2(0\mid t_b)=x_3(0\mid t_b).
\end{align}
Given the environmental variables $\B F$ and $\Lambda_-$, we can formally view $x(a\mid t_b)$ as a function of the environmental variables:
\begin{equation*}
x(a\mid t_b)=\Phi(a, t_b, \B F, \Lambda_-),
\end{equation*}
i.e.\ $x(a\mid t_b)$ is completely determined by
\begin{equation*}
\B F\restrict{[t_b, t_b+a]}, \quad \text{and} \quad \Lambda_-\restrict{[t_b, t_b+a]}.
\end{equation*}

We now define the environmental variables in terms of p-level fractions. Note that $\Lambda_-$ has the exact same interpretation as in network cases I and II. It should therefore come as no surprise that the definition of $\Lambda_-$ in terms of p-level fractions is again~\eqref{def:Lambda-}. The fractions of free binding sites with disease status $d$ are again defined by~\eqref{def:F}. This is \emph{step 2}.

Next, in \emph{step 3}, we define the i-level probabilities $p_{(-,\B k)}(a\mid t_b)$ in terms of $x$. As long as  no infection occurs and the owner does not die, binding sites with the same owner are i.i.d.\ with distribution $x$. Therefore
\begin{equation}\label{def:x2pIII}
p_{(-,\B k)}(a\mid t_b)=\frac{n!}{k_0!\,k_1!\, k_2!\,k_3!}\left(x_0^{k_0}x_1^{k_1}x_2^{k_2}x_3^{k_3}\right)(a\mid t_b)
\end{equation}
(compare with eq.~\eqref{def:x2pII} and note that we condition on the survival of the individual).

In \emph{step 4} we relate p-level fractions $P_{(d, \B k)}(t)$ to i-level probabilities $p_{(d, \B k)}(a\mid t_b)$. In order to do so, we use the stationary age distribution with density $a\mapsto \mu e^{-\mu a}$. The fraction of the population that is in state $(d, \B k)$ at time $t$ is obtained by adding all individuals in that state that are born before time $t$ and are still alive at time $t$. We find that
\begin{align}
P_{(d,\B k)}(t)&=\int_{-\infty}^t \mu e^{-\mu (t-t_b)}p_{(d,\B k)}(t-t_b\mid t_b)dt_b\nonumber\\
&=\int_0^\infty \mu e^{-\mu a}p_{(d,\B k)}(a\mid t-a)da,\label{def:i2pIII}
\end{align}
$d\in\{-,+,\ast\}$.

In \emph{step 5}, we express the environmental variables $\Lambda_-$ and $F_-$ in terms of $x$. This can be done by combining~\eqref{def:x2pIII} and~\eqref{def:i2pIII} with~\eqref{def:Lambda-} (for $\Lambda_-$) or~\eqref{def:F} (for $F_-$). We find that 
\begin{equation}\label{def:Lambda_c}
\Lambda_-(t)=(n-1)\frac{\int_0^\infty\mu e^{-\mu a}x_1x_2\bar x^{n-2}(a\mid t-a)da}{\int_0^\infty\mu e^{-\mu a}x_1\bar x^{n-1}(a\mid t-a)da},
\end{equation}
and
\begin{equation}\label{def:F0_c}
F_-(t)=\int_0^\infty\mu e^{-\mu a}x_0\bar x^{n-1}(a\mid t-a)da.
\end{equation}
In order to complete \emph{step 5} (expressing the environmental variables $F_+$ and $\Lambda_-$ in terms of $x$) we need to consider infectious and recovered binding sites.

\subsubsection{After susceptibility is lost}\label{sec:infIII}
Consider a binding site that was born at time $t_b$ and infected at age $a_+$ and remains alive and infectious for the period under consideration. Note that age $a_+$ for this individual corresponds to calendar time $t_+=t_b+a_+$. Let $y^\text{e}_i(a\mid t_b, a_+)$ denote the probability that the exceptional binding site is in state $i$ at age $a$ and $y_i(a\mid t_b, a_+)$ the same probability for a non-exceptional binding site. 

Then, at age $a_+$, the exceptional binding site is for certain in state 2, while the other $n-1$ binding site states are distributed according to $x(a_+\mid t_b)/\bar x(a_+\mid t_b)$:
\begin{align*}
y_0^\text{e}(a_+\mid t_b, a_+)=0,\qquad y_0(a_+\mid t_b, a_+)&=\frac{x_0}{\bar x}(a_+\mid t_b),\\
y_1^\text{e}(a_+\mid t_b, a_+)=0,\qquad y_1(a_+\mid t_b, a_+)&=\frac{x_1}{\bar x}(a_+\mid t_b),\\
y^\text{e}_2(a_+\mid t_b, a_+)=1,\qquad y_2(a_+\mid t_b, a_+)&=\frac{x_2}{\bar x}(a_+\mid t_b),\\
y^\text{e}_3(a_+\mid t_b, a_+)=0,\qquad y_3(a_+\mid t_b, a_+)&=\frac{x_3}{\bar x}(a_+\mid t_b). 
\end{align*}
The dynamics of infectious binding sites are described by:
\begin{equation}\label{def:ODEyIII}
\begin{aligned}
%\frac{d y^\text{e}(a\mid t_b, a_+)}{d a}&=M_+\big(\B{F}(t_b+a),\Lambda_+(t_b+a)\big)y^\text{e}(a\mid t_b, a_+),\\
\frac{d y(a\mid t_b, a_+)}{d a}&=M_+\big(\B{F}(t_b+a),\Lambda_+(t_b+a)\big)y(a\mid t_b, a_+),
\end{aligned}
\end{equation}
with 
\begin{align*}
M_+(\B{F},\Lambda_+)&=\begin{pmatrix}-\rho F &\sigma+\mu &\sigma+\mu &\sigma+\mu\\\rho F_-& -(\beta\Lambda_++\sigma+\mu) &0&0\\\rho F_+&\beta\Lambda_+&-(\sigma+\mu+\gamma)&0\\\rho F_\ast&0&\gamma&-(\sigma+\mu)\end{pmatrix}.
\end{align*}
Again, there is no rate $\gamma$ in $M_+(\B F, \Lambda_+)$ of leaving the system of infectious binding sites as we assume that infectious binding sites remain infectious in the period under consideration. 

In~\eqref{def:ODEyIII} we can consider $\Lambda_+$ as `known'. Indeed, by combining~\eqref{def:Lambda+} with~\eqref{def:i2pIII} and~\eqref{def:x2pIII}, we can express $\Lambda_+$ in terms of $x$ as follows:
\begin{align*}
\Lambda_+(t)&=1+(n-1)\frac{\int_0^\infty \mu e^{-\mu a}x_2^2\bar x^{n-2}(a\mid t-a)da}{\int_0^\infty\mu e^{-\mu a}x_2\bar x^{n-1}(a\mid t-a)da}.
\end{align*}

We now set out to derive an expression for $F_+$. The probability $\phi_{(+,\B k)}(t_b, a\mid a_+)$ that an individual, born at time $t_b$ and infected at age $a_+$, is in state $(+,\B k)$ at age $a\geq a_+$ is given by
\begin{align}
\phi_{(+,\B k)}(a\mid t_b, a_+)=&\frac{n!}{k_0!\,k_1!\,k_2!\,k_3!}\left(\frac{k_0}n\,y_0^\text{e}\ y_0^{k_0-1}y_1^{k_1}y_2^{k_2}y_3^{k_3}+\frac{k_1}n\,y_1^\text{e}\ y_0^{k_0}y_1^{k_1-1}y_2^{k_2}y_3^{k_3}\right.\nonumber\\
&\quad\left.+ \frac{k_2}n\,y_2^\text{e}\ y_0^{k_0}y_1^{k_1}y_2^{k_2-1}y_3^{k_3}+\frac{k_3}n\, y_3^\text{e}\ y_0^{k_0}y_1^{k_1}y_2^{k_2}y_3^{k_3-1}\right)(a\mid t_b, a_+).\label{def:inf_pc}
\end{align}
The contribution to the incidence of individuals of age $a_+$, born at time $t_b$ and alive for the period under consideration, is given by 
\begin{equation*}
\beta n x_2\bar x^{n-1}(a_+\mid t_b),
\end{equation*}
where the reasoning is similar to cases I and II. Then, taking into account all possible ages of infection $0\leq a_+\leq a$, and the probability that as yet recovery did not occur, the probability that an individual, born at time $t_b$, is in state $(+,\B k)$ at age $a$ is given by
\begin{equation*}
p_{(+,\B k)}(a\mid t_b)=\int_0^a e^{-\gamma(a-a_+)}\beta nx_2\bar x^{n-1}(a_+\mid t_b)\phi_{(+,\B k)}(a\mid t_b, a_+)da_+.
\end{equation*}
The p-level fractions $P_{(+,\B k)}(t)$ at time $t$ are obtained through relation~\eqref{def:i2pIII}. In this way, the dynamics of infectious binding sites describe the dynamics of infectious individuals and the population of such individuals. 

In particular, we find that $F_+$ is defined in terms of infectious (and susceptible) binding sites as follows:
\begin{align*}
F_+(t)=\frac1n\int_0^\infty\mu e^{-\mu a}\int_{0}^a& e^{-\gamma(a-a_+)}\beta n x_2\bar x^{n-1}(a_+\mid t-a)\\
&\left(y_0^\text{e}\, \bar y^{n-1}+(n-1)\bar y^\text{e}\, y_0\, \bar y^{n-2}\right)(a\mid t-a, a_+)da_+da.
\end{align*}
Since $y^\text{e}$ and $y$ are probability vectors, they sum to one, i.e.\ $\bar y^\text{e}(a\mid t_b, a_+)=1=\bar y(a\mid t_b, a_+)$. 
Moreover, with $\varphi_1$ given by~\eqref{eq:epsilon1III}, since
\begin{align}
y_0^\text{e}(a\mid t_b, a_+)&=\varphi_1(a-a_+),\label{eq:y0_excIII}\\
y_0(a\mid t_b, a_+)&=y_0^\text{e}( a\mid t_b, a_+)+\frac{x_0}{\bar x}(a_+\mid t_b)\,e^{-(\rho F+\sigma+\mu)(a-a_+)},\label{eq:y0_III}
\end{align}
we can express $F_+$ in terms of the history of $x$:
\begin{align}
F_+(t)=\frac1n\int_0^\infty&\mu e^{-\mu a}\int_{0}^a e^{-\gamma(a-a_+)}\beta nx_2\bar x^{n-1}( a_+\mid t-a)\Big(\varphi_1(a-a_+)\nonumber\\
&+(n-1)\left(\varphi_1(a-a_+)+\frac{x_0}{\bar x}(a_+\mid t-a)\,e^{-(\rho F+\sigma+\mu)(a-a_+)}\right)\Big)da_+da\label{def:F+c_bs}.
\end{align}

We can use the consistency condition for the total fraction of free binding sites:
\begin{equation}\label{eq:F3III}
F_\ast(t)=F-F_-(t)-F_+(t).
\end{equation}
to express $F_\ast$ in terms of the history of $x$ by using~\eqref{def:F0_c} and~\eqref{def:F+c_bs}. 

Thus we have specified all environmental variables for~\eqref{ODE_III_1} in terms of (the history of) $x$. For completeness we briefly consider recovered binding sites. 

Suppose a recovered binding site was born at time $t_b$, infected at age $a_+$, and recovered at age $a_\ast$ (and as usual, suppose its owner does not die in the period under consideration).We consider probabilities $z_i^\text{e}(a\mid t_b, a_+, a_\ast)$ and $z_i(a\mid t_b, a_+, a_\ast)$ for recovered exceptional and non-exceptional binding sites in state $i$, respectively. The $y$ and $y^\text{e}$ probabilities yield the conditions for $z$ and $z^\text{e}$ at age $a=a_\ast$, i.e.
\begin{alignat*}{3}
z_0^\text{e}(a_\ast\mid t_b, a_+, a_\ast)&=y_1^\text{e}(a_\ast\mid t_b, a_+),\qquad &z_0(a_\ast\mid t_b, a_+, a_\ast)&=y_1(a_\ast\mid t_b, a_+),\\
z_1^\text{e}(a_\ast\mid t_b, a_+, a_\ast)&=y_1^\text{e}(a_\ast\mid t_b, a_+),\qquad &z_1(a_\ast\mid t_b, a_+, a_\ast)&=y_1(a_\ast\mid t_b, a_+),\\
z^\text{e}_2(a_\ast\mid t_b, a_+, a_\ast)&=y_2^\text{e}(a_\ast\mid t_b, a_+),\qquad &z_2(a_\ast\mid t_b, a_+, a_\ast)&=y_2(a_\ast\mid t_b, a_+),\\
z^\text{e}_3(a_\ast\mid t_b, a_+, a_\ast)&=y_3^\text{e}(a_\ast\mid t_b, a_+),\qquad &z_3(a_\ast\mid t_b, a_+, a_\ast)&=y_3(a_\ast\mid t_b, a_+).
\end{alignat*}
The dynamics for $z$ and $z^\text{e}$ can be described by a system of ODE similar to the ODE systems~\eqref{def:ODEyIII} for $y$ and $y^\text{e}$. Only now the mean field at distance one quantity $\Lambda_+$ needs to be replaced by $\Lambda_\ast$ where $\Lambda_\ast$ is defined in terms of p-level fractions by~\eqref{def:Lambda*}. By combining~\eqref{def:Lambda*} with~\eqref{def:i2pIII} and~\eqref{def:x2pIII} we can express $\Lambda_\ast$ in terms of $x$-probabilities:
\begin{equation*}
\Lambda_\ast(t)=(n-1)\frac{\int_0^\infty \mu e^{-\mu a}x_2x_3\bar x^{n-2}(a\mid t-a)da}{\int_0^\infty\mu e^{-\mu a}x_3\bar x^{n-1}(a\mid t-a)da}.
\end{equation*}

Let $\psi_{(\ast, \B k)}(a\mid t_b, a_+, a_\ast)$ denote the probability that a recovered individual is in state $(\ast,\B k)$ given that it was born at time $t_b$, infected at age $a_+$ and recovered at age $a_\ast$, and does not die in the period under consideration. Then $\psi_{(\ast,\B k)}(a\mid t_b, a_+, a_\ast)$ can be expressed in terms of $z$ and $z^\text{e}$ by replacing $\phi$ by $\psi$, $y_i$ by $z_i$, and $y_i^\text{e}$ by $z_i^\text{e}$ in~\eqref{def:inf_pc}.

The probability $p_{(\ast, \B k)}(a\mid t_b)$ is then obtained by taking into account all possibilities for age of infection $a_+$ and age of recovery $a_\ast$:
\begin{equation*}
p_{(\ast,\B k)}(a\mid t_b)=\int_{a_\ast=0}^a\int_{a_+=0}^{a_\ast}\gamma e^{-\gamma(a_\ast-a_+)}\beta nx_2\bar x^{n-1}(a_+\mid t_b)\psi_{(\ast, \B k)}(a\mid t_b, a_+,a_\ast)da_+da_\ast.
\end{equation*}
Finally, by relation~\eqref{def:i2pIII}, we obtain %the fraction $P_{(\ast, \B k)}(t)$, i.e.\ the fraction of the population in state $(\ast,\B k)$ at time $t$.
\begin{equation*}
P_{(\ast, \B k)}(t)=\int_0^\infty\mu e^{-\mu a}p_{(\ast,\B k)}(a\mid t-a)da.
\end{equation*}

\subsubsection{A system of three renewal equations}
To summarize, by replacing $F_\ast$ by~\eqref{eq:F3III}, we are left with three environmental variables $\Lambda_-$, $F_-$, and $F_+$ which are defined by
\begin{align}
\Lambda_-(t)&=(n-1)\frac{\int_0^\infty\mu e^{-\mu a}x_1x_2\bar x^{n-2}(a\mid t-a)da}{\int_0^\infty\mu e^{-\mu a}x_1\bar x^{n-1}(a\mid t-a)da},\label{eq:envLambda}\\
F_-(t)&=\int_0^\infty\mu e^{-\mu a}x_0\bar x^{n-1}(a\mid t-a)da,\label{eq:envF0}\\
F_+(t)&=\frac1n\int_0^\infty \mu e^{-\mu a}\int_{0}^a e^{-\gamma(a-a_+)}\beta nx_2\bar x^{n-1}( a_+\mid t-a)\Big\{\varphi_1(a-a_+)\nonumber\\
&\phantom{=\qquad}+(n-1)\left(\varphi_1(a-a_+)+\frac{x_0}{\bar x}(a_+\mid t-a)\,e^{-(\rho F+\sigma+\mu)(a-a_+)}\right)\Big\}da_+da.\label{eq:envF1}
\end{align}
Recall that $x(a\mid t-a)$ is completely determined by 
\begin{equation*}
F_-\restrict{[t-a, t]},\quad F_+\restrict{[t-a, t]}, \quad \text{and} \quad \Lambda_-\restrict{[t-a, t]},
\end{equation*}
via~\eqref{ODE_III_1}-\eqref{ODE_III_3}. Therefore~\eqref{eq:envLambda}-\eqref{eq:envF1} is a closed system of three renewal equations.

Together, the three renewal equations~\eqref{eq:envLambda}-\eqref{eq:envF1} fully determine the dynamics of i-level probabilities \mbox{$p_{(-,\B k)}(a\mid t_b)$} and p-level fractions $P_{(-,\B k)}(t)$. (Note that there are in total $1/6(n+1)(n+2)(n+3)$ states of the form $(-,\B k)$, with $\B k=(k_1,k_2,k_3)$, $0\leq k_1+k_2+k_3\leq n$.) 

One may not particularly like renewal equations to work with. However, the ODE system~\eqref{ODE_III_1} has $t_b$ as a parameter, so is not finite dimensional. Therefore, contrary to Section~\ref{sec:intermediate}, in order to describe the model with a closed finite system of ODE one needs to turn to p-level fractions $P_{(-,\B k)}$ and $P_{(+,\B k)}$ (the p-level system of ODE can be written down directly from the interpretation; see also~\citep{Leung2015a} and Remarks~\ref{rk:ODEpI} and~\ref{rk:ODEp+I}). Together with the definition of the environmental variables $F_\pm$ and $\Lambda_\pm$ in terms of p-level fractions, the system is then closed. However, there are in total $1/3(n+1)(n+2)(n+3)$ variables of the form $P_{(\pm,\B k)}$.

As the system of three renewal equations~\eqref{eq:envLambda}-\eqref{eq:envF1} has a clear interpretation, and $R_0$, $r$, and the endemic steady state can very nicely be characterized from this system (see Section~\ref{sec:R0III} below), we strongly advocate this formulation of the model rather than a (very high-dimensional) system with only ODE.

\subsection{The beginning of an epidemic: $R_0$ and $r$}\label{sec:R0III}
To describe the beginning of an epidemic, we are interested in characterizing $R_0$ and $r$. We have done so for the full p-level ODE system in~\citep{Leung2015a}. In this paper, the characterization of $R_0$ involved the dynamics of infectious binding sites in the beginning of the epidemic. This infectious binding site system was then, via a linear map, coupled to the linearized p-system to show that the definition of $R_0$ via the interpretation indeed yields a threshold parameter with threshold value one for the p-level system. 

In this section, we use the system of three renewal equations~\eqref{eq:envLambda}-\eqref{eq:envF1} to characterize $R_0$ and $r$. Using the same arguments as in Sections~\ref{sec:R0I} and~\ref{sec:R0II} of network cases I and II, we deduce that, in order to find a threshold parameter for the disease free steady state of the p-level system, we can focus on a threshold parameter for the stability of the disease free steady state of the binding site level system~\eqref{ODE_III_1}. Hence we can focus on~\eqref{eq:envLambda}-\eqref{eq:envF1}.

The linearization of~\eqref{eq:envLambda}-\eqref{eq:envF1} involves the linearization of~\eqref{ODE_III_1}. The disease free steady state of~\eqref{ODE_III_1} is given by $\tilde x_0(a\mid t_b)=\varphi_0(a)$, $\tilde x_1(a\mid t_b)=1-\varphi_0(a)$, $\tilde x_2(a\mid t_b)=0=\tilde x_3(a\mid t_b)$, where $\varphi_0(a)$, the probability that a binding site is free at age $a$ given that it was born free (i.e.\ free at age 0), is given by~\eqref{eq:epsilon0III}. 

We again put a $\wedge$ on the symbols to denote the variables in the linearized system. The ODE for the linearized variable $\hat x_2$ is straightforward:
\begin{align}
\frac{d\hat x_2}{da}(a\mid t_b)&=\rho \hat F_+(t_b+a)\varphi_0(a)+\beta\hat\Lambda_-(t_b+a)\big(1-\varphi_0(a)\big)\nonumber\\
&\phantom{=\ }-(\sigma+\mu+\beta+\gamma)\hat x_2(a\mid t_b),\label{eq:ODEhatx2}\\
\hat x_2(0\mid t_b)&=0.\nonumber
\end{align}
In the following we condition (as usual) on the owner of the binding site staying alive in the period under consideration. The probability $y_0^\text{e}(a\mid t_b,a_+)$ is independent of $t_b$ and given by~\eqref{eq:y0_excIII}. On the other hand, $y_0(a\mid t_b,a_+)$ in the disease free steady state can be interpreted as the probability that a binding site is free at age $a$ given that it is free at age $a_+$ with probability $\varphi_0(a_+)$. But this is equal to the probability $\varphi_0(a)$ that a binding site is free at age $a$ given that it was born free at age 0 (since then, the probability that it is free at age $a_+$ is exactly $\varphi_0(a_+)$). So we find that, in the disease free steady state, 
\begin{equation*}
y_0(a\mid t_b,a_+)=\varphi_1(a-a_+)+e^{-(\rho F+\sigma+\mu)(a-a_+)}\varphi_0(a_+)=\varphi_0(a),
\end{equation*}
where the first equality follows from simply evaluating~\eqref{eq:y0_III} in the disease free steady state and the second can be deduced (as above) from the interpretation (or by algebraic manipulation). So we find that $\hat F_+$ satisfies
\begin{align}
\hat F_+(t)=\frac1n\int_0^\infty\mu e^{-\mu a}\int_{0}^a& e^{-\gamma(a-a_+)}\beta n \hat x_2(a_+\mid t-a)\nonumber\\
&\Big(\varphi_1(a-a_+)\,+(n-1)\varphi_0(a)\Big)da_+da.\label{eq:ODEhatF+}
\end{align}
Next, linearization of $\Lambda_-$ yields
\begin{align}
\hat\Lambda_-(t)&=\frac{1}{1-F}\int_0^\infty \mu e^{-\mu a}(n-1)\big(1-\varphi_0(a)\big)\hat x_2(a\mid t-a)da\label{eq:ODEhatLambda-},
\end{align}
where we used relation~\eqref{eq:Fvarphi_0} between $F$ and $\varphi_0$. 

We now derive two renewal equations for $\hat F_+$ and $\hat \Lambda_-$. Variation of constants yields an expression for $\hat x_2$ in terms of $\hat F_+$ and $\hat \Lambda_-$:
\begin{equation}\label{eq:hatx2}
\hat x_2(a\mid t_b)=\int_0^ae^{-(\sigma+\mu+\beta+\gamma)(a-\alpha)}\left(\rho \hat F_+(t_b+\alpha)\varphi_0(\alpha)+\beta\hat \Lambda_-(t_b+\alpha)\big(1-\varphi_0(\alpha)\big)\right)d\alpha.
\end{equation}
We substitute this in the expressions for $\hat F_+$ and $\hat\Lambda_-$ to find the system of two renewal equations:
\begin{align*}
\hat F_+(t)&=\int_0^\infty\int_0^a\int_0^{a_+}\mu e^{-\mu a}e^{-\gamma(a-a_+)}\beta e^{-(\sigma+\mu+\beta+\gamma)(a_+-\alpha)}\\
&\phantom{=\int\ }\left(\rho \hat F_+(t-a+\alpha)\varphi_0(\alpha)+\beta\hat\Lambda_-(t-a+\alpha)\big(1-\varphi_0(\alpha)\big)\right)\\
&\phantom{=\int\ }\Big(\varphi_1(a-a_+)\,+(n-1)\varphi_0(a)\Big)d\alpha da_+da\\
\hat\Lambda_-(t)&=\frac{1}{1-F}\int_0^\infty\int_0^a \mu e^{-\mu a}(n-1)\big(1-\varphi_0(a)\big)e^{-(\sigma+\mu+\beta+\gamma)(a-\alpha)}\\
&\phantom{=\int\int\int}\left(\rho \hat F_+(t-a+\alpha)\varphi_0(\alpha)+\beta\hat\Lambda_-(t-a+\alpha)\big(1-\varphi_0(\alpha)\big)\right)d\alpha da.
\end{align*}
In preparation for defining and interpreting $R_0$ we write these integrals in convolution form:
\begin{align}
\hat F_+(t)&=\int_0^\infty\int_0^\infty\int_0^\infty\beta e^{-(\sigma+2\mu+\beta+\gamma)\xi}e^{-(\gamma+\mu)\tau}\nonumber\\
&\phantom{=\int\ }\left(\rho F \hat F_+(t-\tau)\pi_0(\alpha)+\beta(1-F)\hat\Lambda_-(t-\tau)\pi_1(\alpha)\right)\nonumber\\
&\phantom{=\int\ }\big(\varphi_1(\tau)\,+(n-1)\varphi_0(\tau+\xi+\alpha)\big) d\alpha d\xi d\tau  \label{eq:hatF+}\\
\hat\Lambda_-(t)&=\frac{1}{1-F}\int_0^\infty\int_0^\infty (n-1)\big(1-\varphi_0(\tau+\alpha)\big)e^{-(\sigma+2\mu+\beta+\gamma)\tau}\nonumber\\
&\phantom{=\int\int\int}\left(\rho F \hat F_+(t-\tau)\pi_0(\alpha)+\beta(1-F)\hat\Lambda_-(t-a+\alpha)\pi_1(\alpha)\right)d\alpha d\tau.\label{eq:hatLambda-}
\end{align}
(the $\pi_0$ and $\pi_1$ appear by multiplying with $F/ F$ and $(1-F)/(1-F)$). This is a system of two renewal equations of the form
\begin{equation}\label{genRE}
\tilde b(t)=\int_0^\infty \tilde K(\tau)\tilde b(t-\tau)d\tau,
\end{equation}
with non-negative kernel $\tilde K$. 

From these two renewal equations~\eqref{eq:hatF+} and~\eqref{eq:hatLambda-}, we can obtain the characteristic equation and deduce threshold parameters $r$ and $R_0$. We define  
\begin{equation}\label{R0_III}
R_0 = \text{dominant eigenvalue of }\int_0^\infty \tilde K(\tau)d\tau.
\end{equation} 
Note that $\int_0^\infty\tilde K(\tau)$ is a $2\times2$ matrix that can be evaluated explicitly so we have an explicit expression for $R_0$. We define $r$ to be the real root (if it exists) of the \emph{characteristic equation}
\begin{equation}\label{r_III}
\det\left(I-\int_0^\infty e^{-\lambda\tau}\tilde K(\tau)d\tau\right)=0
\end{equation}
such that the spectral radius of $\int_0^\infty e^{-\lambda\tau}\tilde K(\tau)d\tau$ equals 1. Note that $r$ is necessarily the rightmost solution of the characteristic equation~\eqref{r_III}. 

Then $r$ is a threshold parameter with threshold value zero for the stability of the disease free steady state of the system of renewal equations~\eqref{eq:envLambda}-\eqref{eq:envF1}. Furthermore sign($R_0-1)=$ sign($r$) so the definition~\eqref{R0_III} of $R_0$ indeed has the right threshold property. 

For $R_0>1$, to see that sign($R_0-1)=$ sign($r$), one uses that each matrix element of $\int_0^\infty e^{-\lambda\tau}K(\tau)d\tau$ is a strictly monotonically decreasing function of $\lambda$ and therefore the dominant eigenvalue of $\int_0^\infty e^{-\lambda\tau}K(\tau)d\tau$ is strictly monotonically decreasing as a function of $\lambda$~\citep{Li2002},\citep[Section 8.2 the intrinsic growth rate]{Diekmann2013}. For $R_0<1$, one uses that the rightmost real solution $r$ of~\eqref{r_III} (if it exists) is strictly less than zero and this establishes the stability of the disease free steady state~\citep{Heijmans1986, Inaba1990, Thieme2009}.

In the epidemic context, `reproduction' corresponds to transmission of the infectious agent to another host. The definition of (and the derivation of an expression for) $R_0$ in~\citep{Leung2015a} is in this spirit: it follows infectious binding sites in time and counts how many new infectious binding sites are formed when transmission occurs. A slight modification of the derivation in~\citep{Leung2015a} is required to generalize from SI to SIR. We did check that~\eqref{R0_III} is identical to the appropriately modified version of the dominant eigenvalue of (59) in Appendix C of~\citep{Leung2015a}.

Yet we would like to have a direct interpretation of the would-be reproduction number~\eqref{R0_III}. To achieve this, it is helpful to think in terms of reproduction `opportunities'. In the present context, these consist of $+-$ links. In~\citep{Leung2015a} the spotlight is on the $+$ side of the link. The present bookkeeping scheme focuses on $x$, so on $-$ binding sites. So now the spotlight is on the $-$ side of the link. The difference is just a matter perspective. A key point, however, is that after transmission the link disappears from the $x$ stage. This forces us to formulate the interpretation in terms of reproduction opportunities rather than reproductions. (Note that, in traditional epidemiological models involving the random mixing assumption, contacts between individuals are instantaneous so there are no $-+$ links or `reproduction opportunities' in the above sense.) 

We distinguish two birth-types of $-+$ links, according to the way they originate:
\begin{description}
\item[Type 0] the $-+$ link was formed when a $-$ binding site and a $+$ binding site linked up
\item[Type 1] the $-+$ link is a transformed $--$ link (one of the two owners got infected by one of its other partners)
\end{description}

\begin{figure}[H]
\centering
\includegraphics[scale=0.55]{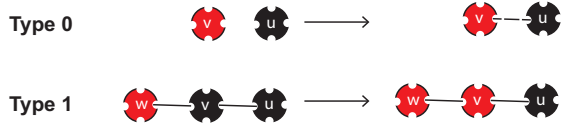}
\caption{The two birth-types of $+-$ links between individuals $u$ and $v$.}
\label{fig:types}
\end{figure}

The relevant difference is the age distribution of the $-$ binding site at the `birth' of the $-+$ link (see Fig.~\ref{fig:types}):
\begin{itemize}
\item for type 0 this distribution has density $\pi_0$ since the $-$ binding site was free until that moment
\item for type 1 this distribution has density $\pi_1$ since the $-$ binding site was (and remains) occupied
\end{itemize}
So the density of the age distribution of the $-$ binding site at birth depends on the birth-type, making it necessary to distinguish between the two birth-types 0 and 1, such in contrast to case II.

\begin{figure}[H]
\centering
\includegraphics[scale=0.6]{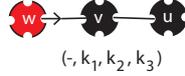}
\caption{An example of a $--+$ configuration: $u$ is one of the $k_1$ $-$ partners of the `middle' $-$ individual $v$ in state $(-,k_1,k_2,k_3)$ and $w$ is one of the $k_2$ $+$ partners of $v$.}
\label{fig:uvw}
\end{figure}

In the nonlinear setting, the total rate in the population at which $-+$ links of type 0 are formed is equal to $\rho F_- \sum k_0 P_{(+,\B k)}=\rho F_- nF_+$ (note the $-+$ asymmetry here, which is in preparation for the linearization). The rate at which type 1 $-+$ links are formed is equal to $\beta \sum k_1 k_2 P_{(-,\B k)}$, respectively. Indeed, the expected number of free infectious binding sites in the population is $\sum k_0 P_{(+,\B k)}$, and the rate at which a free and infectious binding site acquires a susceptible partner is $\rho F_-$. The expected number $--+$ configurations per `middle' $-$ individual is $\sum k_1 k_2 P_{(-,\B k)}$ (see also Fig.~\ref{fig:uvw}) and the rate of transmission is $\beta$. 

Linearization in the disease free steady state yield $\rho F \sum k_0 \hat P_{(+,\B k)}= \rho F n \hat F_+$ and $\beta \sum k_1 \hat P_{(-,k_1,1,0)}=\beta n(1-F)\hat \Lambda_-$, respectively (use p-level definition~\eqref{def:Lambda-} for $\Lambda_-$). These observations motivate us to scale the two renewal equations~\eqref{eq:hatF+} and~\eqref{eq:hatLambda-}. Let
\begin{equation}\label{def:f}
\begin{aligned}
b_0&\coloneqq \rho F n \hat F_+,\\
b_1&\coloneqq \beta n(1-F)\hat \Lambda_-.
\end{aligned}
\end{equation}
That such a rescaling does not affect the definition of $r$ and $R_0$ follows from the following observation:

\begin{Observation}
In general, if we have a system of renewal equations of the form~\eqref{genRE}, we may `scale' $\tilde b$, i.e.\ put $\tilde b_i=c_i b_i$ and consider the renewal equation 
\begin{equation*}
b(t)=\int_0^\infty K(\tau)b(t-\tau)d\tau
\end{equation*}
with $K(\tau)\coloneqq C^{-1}\tilde K(\tau)C$ and $C$ the diagonal matrix with non-zero entries $C_{ii}=c_i$. Then 
\begin{equation*}
\det\left(I-\int_0^\infty e^{-\lambda\tau}\tilde K(\tau)d\tau\right)=\det\left(I-\int_0^\infty e^{-\lambda\tau} K(\tau)d\tau\right).
\end{equation*}
Moreover, the matrices $\int_0^\infty \tilde K(\tau)d\tau$ and $\int_0^\infty K(\tau)d\tau=C^{-1}\int_0^\infty \tilde K(\tau)d\tau C$ are similar, so they have the same eigenvalues. In particular, they have the same dominant eigenvalue $R_0$. 
\end{Observation}

Rescaling~\eqref{def:f} yields a system of renewal equations
\begin{equation}\label{RE_III}
b(t)=\int_0^\infty K(\tau)b(t-\tau)d\tau
\end{equation}
with $b=(b_0\ b_1)$, and $K=(K_{ij})$ a $2\times2$ matrix with matrix elements
\begin{equation}\label{eq:K}
\begin{aligned}
K_{00}(\tau)&= \int_0^\infty\int_0^\infty\pi_0(\alpha)\beta e^{-(\sigma+2\mu+\beta+\gamma)\xi}e^{-(\mu+\gamma)\tau}\\
&\phantom{\int_0^\infty\int_0^\infty}\rho F\big(\varphi_1(\tau)+(n-1)\varphi_0(\tau+\xi+\alpha)\big) d\alpha d\xi\\
K_{01}(\tau)&=\int_0^\infty\int_0^\infty\pi_1(\alpha)\beta e^{-(\sigma+2\mu+\beta+\gamma)\xi}e^{-(\mu+\gamma)\tau} \\
&\phantom{\int_0^\infty\int_0^\infty}\rho F\big(\varphi_1(\tau)+(n-1)\varphi_0(\tau+\xi+\alpha)\big) d\alpha d\xi\\
K_{10}(\tau)&=\int_0^\infty\pi_0(\alpha)e^{-(\sigma+2\mu+\beta+\gamma)\tau}\beta(n-1)\big(1-\varphi_0(\tau+\alpha)\big)d\alpha\\
K_{11}(\tau)&=\int_0^\infty\pi_1(\alpha)e^{-(\sigma+2\mu+\beta+\gamma)\tau}\beta(n-1)\big(1-\varphi_0(\tau+\alpha)\big)d\alpha.
\end{aligned}
\end{equation}
(Again we note that each of these four integrals can be evaluated explicitly.)

We now explain how~\eqref{eq:K} can be interpreted in terms of reproduction opportunities of types 0 and 1. A $-+$ link has no `descendants' when transmission does not occur. When transmission occurs, it has at that very moment descendants of type 1, because the `other' partners of the owner $u$ of the $-$ link then all of a sudden are connected to a $+$ individual. In addition, it has descendants of type 0 when empty binding sites of $u$ get occupied (necessarily by a $-$ partner, since we consider the initial phase when $+$ individuals are rare). Note that we should follow all binding sites of $u$ until $u$ either dies or becomes removed, since occupied binding sites may become free, occupied again, etcetera.

We now compute the expected number of descendants of either type for a $-+$ link given that the owner $u$ of the $-$ binding site has age $a$ at the birth of the $-+$ link. The force of infection on u along the link equals $\beta$ as long as
\begin{itemize}
\item the $+$ partner is alive and infectious
\item separation did not occur
\item $u$ is alive and not yet infected
\end{itemize}
Hence the probability per unit of time that $u$ is infected at age $\alpha+\tau$ is given by $\beta e^{-(\mu+\gamma+\sigma+\mu+\beta)\tau}$. 

\begin{figure}[H]
\centering
\includegraphics[scale=0.55]{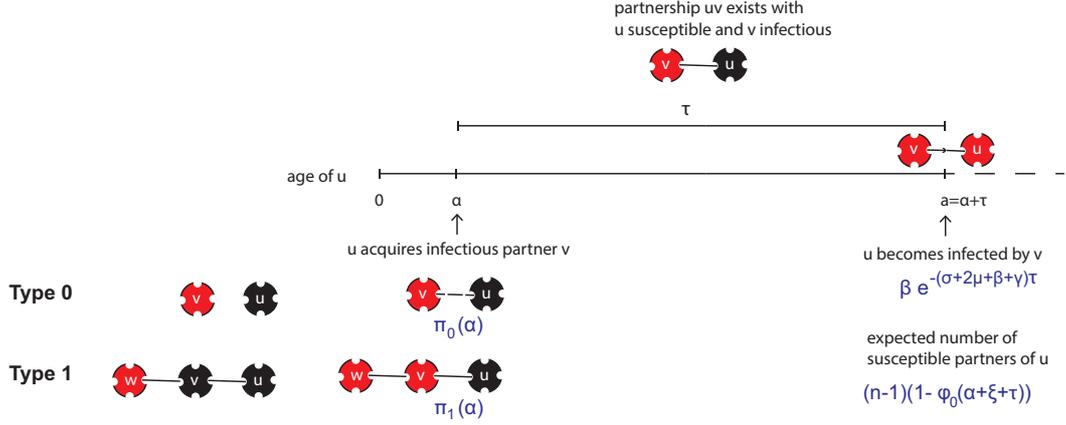}
\caption{The production of type 1 offspring.}
\label{fig:M1}
\end{figure}

When $u$ is infected at age $\alpha +\tau$ an expected number $(n-1)\big(1- \varphi_0(\alpha+\tau)\big)$ of offspring of type 1 is produced. A schematic representation is given in Fig.~\ref{fig:M1}. This is how $\int_0^\infty K_{01}(\tau)d\tau$ and $\int_0^\infty K_{11}(\tau)d\tau$ in~\eqref{eq:K} can be interpreted.

\begin{figure}[H]
\centering
\includegraphics[scale=0.55]{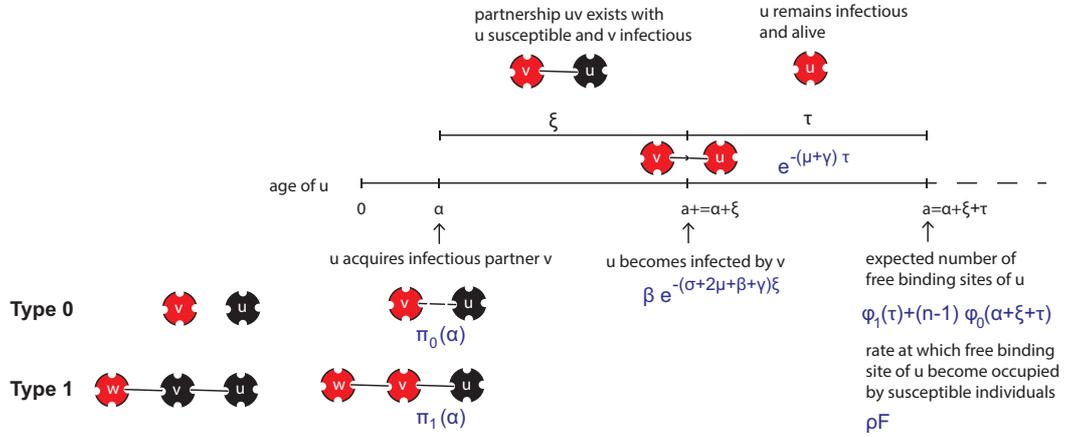}
\caption{The production of type 0 offspring.}
\label{fig:M0}
\end{figure}

Offspring of type 0 is (potentially) produced by both free and occupied (at the time of infection) binding sites of $u$. To calculate the mean number of offspring of type 0, suppose $u$ is infected at age $\alpha+\xi$. Then $u$ is alive and infectious at age $\alpha+\xi+\tau$ with probability $e^{-(\gamma+\mu)\tau}$. The expected number of free binding sites it has at age $\alpha+\tau+\xi$ is equal to $\varphi_1(\xi)\,+(n-1)\varphi_0(\tau+\xi+\alpha)$. A free binding site becomes occupied at rate $\rho F$. Integrating over all possible $\tau>0$, we find that the expected offspring of type 0 is 
\begin{equation*}
\int_0^\infty e^{-(\gamma+\mu)\tau}\rho F\big(\varphi_1(\tau)\,+(n-1)\varphi_0(\tau+\xi+\alpha)\big) d\tau.
\end{equation*}
This is how $\int_0^\infty K_{00}(\tau)d\tau$ and $\int_0^\infty K_{01}(\tau)d\tau$ in~\eqref{eq:K} can be interpreted. A schematic representation is given in Fig.~\ref{fig:M0}.

\paragraph{The infectious binding site perspective}\mbox{}\\
In~\citet{Leung2015a} the focus was on infectious binding sites. As exhibited by in the densities $\pi_0$ and $\pi_1$ of the age-distribution of $-$ individuals in the birth-types of $-+$ links, it matters whether a newly created $-+$ links is type 0 or type 1. To take this into account, in~\citep{Leung2015a}, we kept track of the number of partners of susceptible partners of infectious binding sites. This led to the reduction to a $2\times2$ next-generation-matrix involving mean times spent with a susceptible partner with $k$ partners, $k=1,\ldots,n$, (in the form of the inverse of an $n\times n$ matrix). We were able to find an explicit expression for $R_0$ although it required quite a lot of work to deal with this $n\times n$ inverse matrix.
 
The results in this paper teach us that, to take into account the birth-types of $-+$ links, we can also keep track of the age of susceptible partners rather than partners of partners. While age can be anything from zero to infinity, it can only move forward in time, i.e.\ individuals can only grow older. The same $2\times2$ NGM is obtained in a much more straightforward manner.

Whether doing the bookkeeping of partners of partners or of the age of partners, a big downside of taking the infectious binding site perspective is that it takes quite some work to \emph{prove} that so-obtained $R_0$ is actually a threshold parameter for the stability of the disease free steady state of the p-level system (see~\citep{Leung2015a}). This comes almost for free when taking the $-$ perspective as we did in this paper.

Finally note that, whether we consider actual `reproductions' (taking the $+$ perspective) or `reproduction opportunities' (taking the $-$ perspective), both yield the exact same threshold parameter $R_0$ so in that sense it does not matter which perspective we take. However, while the \emph{the dominant eigenvalue} $R_0$ of the next-generation-matrix is the same with both perspectives, the \emph{matrices} themselves are different. And so are the underlying interpretations. 

\subsection{Endemic steady state}\label{sec:EEIII}
Let $E=(F_+, F_\ast,\Lambda_-)$ be the vector of environmental variables. Note that we use consistency relation~\eqref{eq:F3III} to substitute environmental variable $F_\ast$ for $F_-$. This choice of environmental variables leads to the disease free steady state corresponding to $E=(0,0,0)$. Then we have a system of three renewal equations for $E$. 

Let
\begin{align*}
G_1(E)(t)&\coloneqq\int_0^\infty\mu e^{-\mu a}\int_{0}^a e^{-\gamma(a-a_+)}\beta x_2(a_+\mid t-a)\\
&\qquad\Big(n\frac{\sigma+\mu}{\rho F+\sigma+\mu}\left(1-e^{-(\rho F+\sigma+\mu)(a-a_+)}\right)\bar x(a_+\mid t-a)^{n-1}\\
&\qquad+(n-1)e^{-(\rho F+\sigma+\mu)(a-a_+)}x_0( a_+\mid t-a)\bar x(a_+\mid t-a)^{n-2}\Big)da_+da,\\
G_2(E)(t)&\coloneqq F-\int_0^\infty\mu e^{-\mu a}x_0\bar x^{n-1}(a\mid t-a)da-G_1(E),\\
G_3(E)(t)&\coloneqq(n-1)\frac{\int_0^\infty\mu e^{-\mu a}\left(x_1x_2\bar x^{n-2}\right)(a\mid t-a)da}{\int_0^\infty\mu e^{-\mu a}\left(x_1\bar x^{n-1}\right)(a\mid t-a)da}.
\end{align*}
where $x(a\mid t-a)$ is completely determined by $E\restrict{[t-a,a]}$ via~\eqref{ODE_III_1}-\eqref{ODE_III_3}. Therefore, 
\begin{equation}\label{fixedpoint}
\begin{pmatrix}F_+\\F_\ast\\\Lambda_-\end{pmatrix}=\begin{pmatrix}G_1(F_+, F_\ast,\Lambda_-)\\G_2(F_+, F_\ast,\Lambda_-)\\G_3(F_+, F_\ast,\Lambda_-)\end{pmatrix}
\end{equation}
is a closed system of three renewal equations. 

In endemic equilibrium, the environmental variable $E$ is constant (note that, if $E$ is constant, then also p-level fractions are constant and binding-site- and i-level probabilities are constant as functions of time of birth $t_b$). So the endemic steady state is characterized as a solution to the fixed point problem~\eqref{fixedpoint} where now the symbols denote the values of constant functions. The fixed point problem always has a trivial solution given by the disease free steady state $E=(0,0,0)$. Note that a solution $E$ to~\eqref{fixedpoint} needs to have biological meaning. Therefore, we only consider solutions that satisfy $F_+, F_\ast\geq0$, $0\leq F_++F_\ast\leq F$, and $0\leq \Lambda_-\leq n-1$.

\paragraph{Conjecture:}
If $R_0<1$, then the only solution to the fixed point problem is the trivial solution. If $R_0>1$, then there is a unique nontrivial solution.
\paragraph{Open problem:}
Prove (or disprove) the conjecture.
\mbox{}\\\\
In Appendix~\ref{app:proof} we elaborate on an unsuccessful attempt at a proof of the conjecture for the simpler case of an SI infection, rather than an SIR infection, obtained by setting $\gamma=0$. This attempt tried to use Krasnoselskii's method~\cite{Kras1964} (see also~\cite{Hethcote1985}). 

Note that the three-dimensional fixed point problem~\eqref{fixedpoint} provides a way to find the endemic steady state numerically. Furthermore, even though we did not manage to prove the conjecture, numerical investigations strongly suggest that all conditions for Krasnoselskii's method are satisfied and that the conjecture holds true.

\section{Conclusions and discussion}\label{sec:discussion}
In this paper we formulated binding site models for the spread of infection on networks. The binding sites serve as building blocks for individuals. In fact we considered three different levels: (1) binding sites, (2) individuals, and (3) the population. On both the binding site and individual level, we have a Markov chain description of the dynamics, where feedback from the population is captured by environmental variables. These environmental variables are population-level quantities. By lifting the individual level to the population level (where the model is deterministic), the feedback loop can be closed. In the end, this leads to a model description in terms of susceptible binding sites in case I and in terms of just environmental variables in cases II and III.

The systematic model formulation leads, in all three cases, to only a few equations that determine the binding site, individual, and the population dynamics. Moreover, from these equations we derive the epidemiological quantities of interest, i.e.\ $R_0$, $r$, the final size (in cases I and II) and the endemic steady state (in case III). 

Quite a general understanding is enhanced by an elaboration of the interpretation of $R_0$ in a specific context. In cases I and II we have taken the obvious perspective of a $+$ binding site to do so. But in case III, cf.\ Section~\ref{sec:R0III}, we reasoned in terms of `reproduction opportunities'. These consist of $+-$ links. From these links we took the $-$ perspective. Somewhat surprisingly, this turned out to lead quickly and efficiently to a simple interpretation. Moreover, the derivation of $R_0$ follows from the system of equations in a natural manner. One can adopt the $-$ perspective in cases I and II too, but there it does not change much. Yet we wouldn't be surprised if the $-$ perspective turns out to be powerful in other dynamic network models of infectious disease transmission. 

%In cases I and II we have taken the $+$ binding site perspective to interpret $R_0$. Exactly as one normally does. In case III, cf.\ Section~\ref{sec:R0III}, we reasoned in terms of reproduction opportunities that are formed by $+-$ links and we took the $-$ perspective. This same reasoning from the $-$ perspective rather than the $+$ perspective can also be applied to cases I and II. In fact, it can really pay off to do so, as case III clearly illustrates.

Several open problems remain. Although we are able to implicitly characterize the final size in case II, we have not been able to make it more explicit. We would like a characterization in the same spirit as~\eqref{finalRE} for case I, but we have not succeeded and our optimism subsided. A more useful characterization of the endemic steady state was given for case III as a three-dimensional fixed point problem. Unfortunately, we have not (yet) been able to prove the existence and uniqueness of a nontrivial fixed point for $R_0>1$ (and that no such fixed point exists for $R_0<1$) and therefore we posed this as a conjecture in Section~\ref{sec:EEIII}. 

Of another nature are open problems related to the mean field at distance one assumption. While, in case I, the mean field at distance one assumption is proven to be exact in the appropriate large population limit of a stochastic SIR epidemic on a configuration network, it remains an open problem whether or not this also holds for the dynamic network case II (we conjecture it does). In the dynamic network case III, we know that the mean field at distance one assumption is really an approximation of the true dynamics as we pointed out in the introduction of this paper. What we have not discussed is how good or bad of an approximation it is. In particular, are there conditions for which the approximation works nicely and can we understand intuitively the extent to which this assumption violates the truth? 

In both cases II and III, we ended the model formulation with renewal equations. In case II one can just as easily consider a system of ODE, and we represented this view also in the section title~\ref{sec:REII}. In case III, a system of ODE clearly becomes inconvenient. An ODE formulation in that case would require at least $1/3(n+1)(n+2)(n+3)$ variables, while, by considering a system of renewal equations, only three equations are needed. More importantly, the system of renewal equations has the huge advantage that $R_0$ and $r$ more or less immediately follow from the linearization of the system in the disease free steady state. The calculations are straightforward, the expressions are interpretable biologically, and the proof that $R_0$ and $r$ are threshold parameters for the disease free steady state of the p-level system comes more or less for free.

By distinguishing the three different levels, and formulating the model on the binding site level, one can easily consider several generalizations (see also~\citet{Leung2012,Leung2015a} for a discussion). In principle, any generalization that maintains the (conditional) independence assumption for binding sites of an individual easily fits within this framework. One can think of generalizations concerning the network or generalizations concerning the infectious disease. For the infectious disease, one can easily take any compartmental model such as SIR, SEIR, SI, SI$_1$I$_2$ (as long as infected individuals can not return to the susceptible class within their lifetime). The main difference is in the different states that a binding site can be in. Generalizations of the network that one can think of are (i) a heterosexual population rather than a homosexual population, (ii) allowing for different $n$ in the population, i.e.\ letting $n$ be a random variable (which we already considered in the static network case in Section~\ref{app:RE}) (iii) allowing for multiple types of binding sites, e.g.\ binding sites for casual and steady partnerships, and combinations of the three. One can formulate models incorporating these generalizations by following the five steps described in Section~\ref{sec:steps}. The main added difficulty is in the bookkeeping that becomes more involved. But in terms of the characterization of $R_0$ and the endemic steady state, mathematically speaking the situation does not become more complex. 

Finally, in the current framework, and as usual in literature, demographic turnover as considered in case III takes the individual's age to be exponentially distributed. This assumption is mainly for mathematical convenience and is not realistic for many populations. We believe that it is possible to relax the assumption on the age distribution to consider more general survival functions. In that case, lifting the i-level to the p-level changes, and one needs to take into account the age of partners (but hopefully this may be done by simply averaging in the right way). Moreover, in the current framework, disease does not impact mortality. In the context of HIV, disease-related mortality is certainly very relevant. We believe that the framework presented in this paper provides a way to incorporate this by means of the infectious $y$ binding sites. While these generalizations relating to the demographic process are less straightforward to implement than the ones described in the previous paragraph, the current framework provides an excellent starting point. 

%Both generalizations related to the demographic process seem promising, but more research is needed to investigate the possibilities.  

\setlength{\bibsep}{2pt}
\bibliographystyle{plainnat}
\bibliography{refs_network}

\appendix
\section{Do `far past' conditions single out a unique solution?}\label{initial}
In this paper we duck the responsibility of rigorously showing that the systems that we introduce have, modulo translation, at most one epidemiologically relevant (i.e.\ positive) solution. The aim of this appendix is to sketch  the underlying ideas and to provide some references. 

Linearization of an epidemic system in the disease free steady state leads to a linear system that leaves a cone, characterized by positivity, invariant. Perron-Frobenius theory, or its infinite dimensional Krein-Rutman variant, yields the existence of a simple eigenvalue $r$ such that 
\begin{enumerate}
\item[(i)] the corresponding eigenvector is positive
\item[(ii)] Re $\lambda<r$ for all eigenvalues $\lambda\neq r$
\end{enumerate}
The theory of stable and unstable manifolds yields a nonlinear analogue: the nonlinear system has exactly one orbit that is tangent to the eigenvector corresponding to eigenvalue $r$. If $r>0$ then this orbit belongs to the unstable manifold and tends to the disease free steady state for $t\to-\infty$. If $r<0$ then the orbit belongs to the stable manifold and tends to the disease free steady state for $t\to+\infty$. Our interest is in the case $r>0$. 

Note that one orbit of an autonomous dynamical system corresponds to a family of solutions that are translates of each other. See~\citep{Diekmann1977} for an early example of this type of result (but note that the proof in that paper has a flaw; see~\citep[Section 7]{Diekmann1984} for a flawless proof).

These ideas apply directly to the three-dimensional ODE system~\eqref{eq:ODExI_final} in case I. For the scalar renewal equation~\eqref{RE} we can refer to Section 7 of~\citep{Diekmann1984} provided that we are willing to assume that $\mathcal F'$ has compact support. For the ODE system of case II there exists an eigenvalue zero (corresponding to conservation of binding sites). This eigenvalue zero creates havoc. Presumably, the difficulties can be overcome by the introduction of a tailor-made cone, but we did not elaborate this in all required detail. The alternative is to consider the scalar renewal equation~\eqref{def:F+b_bs} for $F_+$ and to combine ideas from~\citep{Diekmann2007} with theory developed in~\citep{Diekmann2012}. This combination should, we think, also cover the system of renewal equations~\eqref{eq:envLambda}-\eqref{eq:envF1} for case III.

\section{Endemic steady state: unsuccessful attempt at a proof}\label{app:proof}
\begin{figure}[H]
\centering
\includegraphics[scale=0.3]{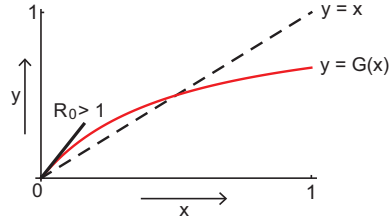}
\caption{Krasnoselskii's method generalizes the geometric arguments in one dimension to multiple dimensions.}
\label{fig:kras}
\end{figure}

We explain our attempt to prove the conjecture of Section~\ref{sec:EEIII} about the existence and uniqueness of solutions to the fixed point problem~\eqref{fixedpoint} for the simpler case of an SI infection rather than an SIR infection (set $\gamma=0$). We only need to consider two environmental variables, rather than three, as we will explain. This attempt to prove the conjecture uses the sublinearity  method of~\cite{Kras1964} (see also~\cite{Hethcote1985}), the idea of which for one dimension is represented in Fig.~\ref{fig:kras}. 

First of all, if $\gamma=0$, then $x$ satisfies
\begin{equation}\label{SIeq:dynamic2ODEx}
\begin{aligned}
\frac{d x_0}{d a}(a\mid t_b)&=-\rho F x_0(a\mid t_b)+(\sigma+\mu)(x_1(a\mid t_b)+x_2(a\mid t_b))\\
\frac{dx_1}{d a}(a\mid t_b)&=\rho (F-F_+)(t_b+a)x_0(a\mid t_b)-(\sigma+\mu+\beta \Lambda_-(t_b+a))x_1(a\mid t_b)\\
\frac{d x_2}{d a}(a\mid t_b)&=\rho F_+(t_b+a)x_0(a\mid t_b)+\beta \Lambda_-(t_b+a)x_1(a\mid t_b)-(\sigma+\mu+\beta)x_2(a\mid t_b)
\end{aligned}
\end{equation}
with boundary condition
\begin{align}\label{SIeq:dynamic2initial}
x_0(0\mid t_b)&=1,\quad x_1(0\mid t_b)=0=x_2(0\mid t_b).
\end{align}
As before in Section~\ref{sec:demography}, $x(a\mid t_b)$ is completely determined by 
\begin{equation*}
F_+\restrict{[t_b, t_b+a]}, \text{\quad and\quad}\Lambda_-\restrict{[t_b, t_b+a]},
\end{equation*}
via~\eqref{SIeq:dynamic2ODEx}-\eqref{SIeq:dynamic2initial}. In particular, there are now only two environmental variables $F_+$ and $\Lambda_-$. These environmental variables satisfy renewal equations. Let
\begin{align*}
G_1(F_+, \Lambda_-)(t)&=F-\int_0^\infty\mu e^{-\mu a}x_0\bar x^{n-1}(a\mid t-a)da,\\
G_2(F_+, \Lambda_-)(t)&=(n-1)\frac{\int_0^\infty\mu e^{-\mu a}x_1x_2\bar x^{n-2}(a\mid t-a)da}{\int_0^\infty\mu e^{-\mu a}x_1\bar x^{n-1}(a\mid t-a)da},
\end{align*}
then we obtain a fixed point problem for the environmental variables $F_+$ and $\Lambda_-$:
\begin{equation}\label{SIfixedpoint}
\begin{pmatrix}F_+\\\Lambda_-\end{pmatrix}=\begin{pmatrix}G_1(F_+,\Lambda_-)\\G_2(F_+,\Lambda_-)\end{pmatrix}.
\end{equation}
Note that in endemic equilibrium the environment is constant, i.e.\ $F_+(t)=\bar F_+$, $\Lambda_-(t)=\bar\Lambda_-$. Therefore $x$ no longer depends on time of birth. In what follows we write $x=x(a)$. 

The fixed point problem~\eqref{SIfixedpoint} can be related to $R_0$ by considering the linearizaton of the right hand side of~\eqref{SIfixedpoint} in the disease free steady state $(F_+,\Lambda_-)=(0,0)$. Indeed, the linearization $DG(0,0)$ has dominant eigenvalue $R_0$. 

Next, Krasnoselskii's method uses the monotonicity of $G_1$ and $G_2$ in both variables $F_+$ and $\Lambda_-$ and strict sublinearity for both $G_1$ and $G_2$, i.e.\ $G_i(t(F_+,\Lambda_-))>tG_i(F_+,\Lambda_-)$ for all $0<t<1$. $i=1,2$). 

Monotonicity and sublinearity of $G_1$ in both variables $F_+$ and $\Lambda_-$ is easily proven. One can show that the derivatives of $x_0$, $x_1$, and $x_1+x_2$ with respect to $F_+$ and $\Lambda_-$ are nonpositive while the mixed second order derivatives are all nonnegative. Then one can easily prove that the derivatives $D_iG_1(F_+,\Lambda_-)\geq0$ showing that $G_1$ is a monotonically increasing function of both $F_+$ and $\Lambda_-$. Sublinearity can be proven by showing that the function $f(t)=G_1(t(F_+,\Lambda_-))-tG_1(F_+,\Lambda_-)$ satisfies $f''(t)<0$ for $0<t<1$. 

We work out only the proof to show that $D_1G_1(F_+,\Lambda_-)\geq0$. The derivative of $G_1$ with respect to $F_+$ is equal to
\begin{equation}\label{F+PDE}
D_1G_1(F_+,\Lambda_-)=-\int_0^\infty\mu e^{-\mu a}\left(\frac{\partial d x_0}{\partial F_+}\bar x^{n-1}+(n-1)\frac{\partial \bar x}{\partial F_+}x_0\bar x^{n-2}\right)(a)da.
\end{equation}
Here $\partial x/\partial F_+$ satisfies:
\begin{equation}\label{PDE}
\begin{aligned}
\frac{d}{da}\frac{\partial x}{\partial F_+}&=M(F_+,\Lambda_-)\frac{\partial x}{\partial F_+}+A_1x\\
\frac{\partial x}{\partial F_+}(0)&=0,
\end{aligned}
\end{equation}
with 
\begin{equation*}
M(F_+,\Lambda_-)=\begin{pmatrix}-\rho F & \sigma+\mu & \sigma+\mu\\\rho (F-F_+)&-(\sigma+\mu+\beta\Lambda_-)&0\\\rho F_+ & \beta\Lambda_-&-(\sigma+\mu+\beta)\end{pmatrix}
\end{equation*}
and
\begin{equation*}
A_1=\rho\begin{pmatrix}0 & 0 & 0\\-1&0&0\\1 & 0&0\end{pmatrix}.
\end{equation*}
To prove that $x_0$, $x_1$, and $x_1+x_2$ are monotonically decreasing functions of $F_+$, we prove that the derivatives with respect to $F_+$ are nonpositive. Working out~\eqref{PDE} we find that
\begin{align*}
\frac{d}{da}\frac{\partial x_0}{\partial F_+}&=-\rho F \frac{\partial x_0}{\partial F_+}+(\sigma+\mu)\left(\frac{\partial x_1}{\partial F_+}+\frac{\partial x_2}{\partial F_+}\right)\\
\frac{d}{da}\frac{\partial x_1}{\partial F_+}&=\rho (F-F_+)\frac{\partial x_0}{\partial F_+}-(\sigma+\mu+\beta\Lambda_-)\frac{\partial x_1}{\partial F_+}-\rho x_0\\
\frac{d}{da}\left(\frac{\partial x_1}{\partial F_+}+\frac{\partial x_2}{\partial F_+}\right)&=\rho F \frac{\partial x_0}{\partial F_+}-(\sigma+\mu+\beta)\left(\frac{\partial x_1}{\partial F_+}+\frac{\partial x_2}{\partial F_+}\right)+\beta \frac{\partial x_1}{\partial F_+},
\end{align*}
where $x_0\geq0$. All off-diagonal terms and the inhomogeneous term are $\leq0$, and the initial conditions for $\partial x_0/{\partial F_+}$, $\partial x_1/{\partial F_+}$, and $\partial x_1/{\partial F_+}+\partial x_2/{\partial F_+}$ are equal to zero. Therefore we find that $\partial x_0/{\partial F_+}(a),\partial x_1/{\partial F_+}(a),\partial x_1/{\partial F_+}(a)+\partial x_2/{\partial F_+}(a)\leq0$ for all $a$, and also $\partial \bar x/{\partial F_+}(a)\leq0$. Similarly, if we replace $F_+$ by $\Lambda_-$ in the partial derivative and $-\rho x_0$ by $-\beta x_1$ then we also find that $x_0$, $x_1$, and $x_1+x_2$ are monotonically decreasing functions of $\Lambda_-$. Together with~\eqref{F+PDE} this shows that $G_1$ is monotonically increasing in both $F_+$ and $\Lambda_-$, i.e.\ $D_1G_1(F_+,\Lambda_-)\leq0$ and $D_2G_1(F_+,\Lambda_-)\leq0$.

\begin{remark}
The variable $x_2$ is not necessarily monotone in $F_+$ or $\Lambda_-$. One can find parameter values for which we find that $\partial x_2/{\partial E}(a)$, $E=F_+,\Lambda_-$, is neither nonpositive nor nonnegative as a function of $a$.
\end{remark}

Note that the feedback function $G_2$ for $\Lambda_-$ involves $x_2$. The arguments to prove monotonicity and sublinearity do not seem to work for $G_2$. Numerical investigation strongly suggest that $G_2$ is indeed monotonically increasing as a function of both $F_+$ and $\Lambda_-$ as well as sublinear. So far, we have not been able to provide a proof.

Nevertheless, once we show that both $G_1$ and $G_2$ are monotonically increasing functions of environmental variables $F_+$ and $\Lambda_-$ and sublinear, Krasnoselskii's method then provides a proof that for $R_0<1$ only the trivial solution exists and for $R_0>1$ there exists a unique nontrivial solution to~\eqref{SIfixedpoint}.

\end{document}